\documentclass[published]{JHEP3} 

\JHEP{00(2007)000}

\JHEPspecialurl{http://jhep.sissa.it/JOURNAL/JHEP3.tar.gz}

\usepackage{epsfig,multicol,bbm}

\newcommand\fverb{\setbox\fverbbox=\hbox\bgroup\verb}
\newcommand\fverbdo{\egroup\medskip\noindent%
            \fbox{\unhbox\fverbbox}\ }
\newcommand\fverbit{\egroup\item[\fbox{\unhbox\fverbbox}]}
\newbox\fverbbox

\title{CMB anisotropies induced by tensor modes in Massive Gravity}

\author{Dennis Bessada$^{1,}$ $^2$, Oswaldo D. Miranda$^2$\\1 Dept. of Physics, The State University of
New York at Buffalo\\239 Fronczak Hall, Buffalo, NY 14260-1500 USA
\\2 INPE - Instituto Nacional de Pesquisas Espaciais - Divis\~ao
de Astrof\'isica, \\Av.dos Astronautas 1758, S\~ao Jos\'e dos
Campos, 12227-010 SP,
 Brazil\\E-mail: \email{dbessada@buffalo.edu}, \email{oswaldo@das.inpe.br}}

\received{\today}       
\accepted{\today}       

\preprint{\hepth{9912999}}  

\abstract{We study Gravitational Waves (GWs) in the context of
Massive Gravity, an extension to General Relativity (GR) where the
fluctuations of the metric have a nonzero mass, and specifically
investigate the effect of the tensor modes on the Cosmic Microwave
Background (CMB) anisotropies. We first study the time evolution
of the tensor modes in Massive Gravity and show that there is a
graviton mass limit $m_l=10^{-66}g\sim 10^{-29}cm^{-1}$, so that
for masses $m\leq m_l$ the tensor perturbations in Massive Gravity
are indistinguishable from the corresponding ones in GR. Also, we
show that short wavelength massive modes behave almost
indistinguishably from their massless counterparts. Later on, we
show that massive gravitons with masses within the range $m=
10^{-27}cm^{-1}$ - $m=10^{-26}cm^{-1}$ would leave a clear
signature on the lower multipoles ($\ell< 30$) in the CMB
anisotropy power spectrum. Hence, our results indicate that CMB
anisotropies measurements might be decisive to show whether the
tensor modes are massive or not.}

\keywords{cosmology of theories beyond the SM, gravity waves /
theory, CMBR theory}

\begin{document}


\section{\label{sec:intr}Introduction and Summary}

Over the last years we have been witnessing a great evolution in
the observation of the universe. An accurate statistical analysis
of the anisotropies of the CMB using the five-year WMAP data
\cite{spergel2008} has shown that the most favored cosmological
model to fit the data is the flat $\Lambda$CDM model, which
includes not only the very well known baryonic matter, but the
mysterious dark matter (DM) and dark energy (DE) as well. The
introduction of dark matter does not require a modification of GR
(despite some alternative models do); in the pure
general-relativistic case, all we have to do is to add further
terms to the energy-momentum tensor on the right-hand side of the
GR field equations. The DM candidates must pass some important
tests \cite{taoso2008} to ensure that the resulting model is
physically consistent. However, despite the good theoretical
candidates available in \cite{taoso2008}, there is no
observational evidence so far to support any of them as the actual
components of DM.

However, the inclusion of DE is not that simple. Ordinary matter,
either baryonic or dark, cannot accelerate the universe as shown
by observations \cite{riess1998}, \cite{perlmutter1999}. This
discovery posed one of the deepest questions in modern cosmology:
is the universe accelerating due to a repulsive gravity (caused by
the quantum energy of the vacuum, for example, as in the
$\Lambda$CDM model), or does General Relativity (GR) break down on
cosmological scales \cite{frieman2008}? Many attempts have been
made addressing these two possible cases, either by introducing
new features into GR, or by modifying it (see \cite{copeland2006}
for a review of the models for DE).

The two key points introduced above are just to illustrate the
need of studying alternative theories of gravity in parallel with
the improvement of models in GR itself. As alternative theories we
mean modifications of GR, and the simplest possibility for
modifying GR is the introduction of a mass for the graviton. It
was performed for the first time by the pioneering work of M.
Fierz and W. Pauli \cite{fierz1939}, where they considered a
linearized field theory of spin-two massive particles. The Lorentz
invariance of the Fierz-Pauli (FP) lagrangian yields a spin-two
massive state with five polarization modes (states with helicities
$\pm 2$, $\pm 1$ and $0$), differing from GR where one finds only
a spin-two state with the two tensor polarization modes
(helicities $\pm 2$). Such extra degrees of freedom yield an
additional contribution of one vector and one real scalar massless
particles with helicities $\pm 1$ and $0$, respectively. The
scalar particle couples to the trace of the stress energy-momentum
tensor, causing a discontinuity in the propagator when one
switches from the massive to the massless regime. This is the
so-called van Dam-Veltman-Zakharov (vDVZ) discontinuity
\cite{veltman1970}, \cite{zakharov1970}, whose net effect for a
theory of a massive spin-two graviton is catastrophic: it would
not even pass the solar-system tests for a theory of gravity (the
prediction of the angle concerning the bending of the light by the
Sun, for example).

However, in a full theory of gravity, we must consider nonlinear
effects; the FP theory is valid only in the linear approximation.
Nonlinear effects eliminate the vDVZ discontinuity in the
classical level \cite{vainshtein1972}, \cite{deffayet2002}, so
that classically we may reconcile the massive theory with the GR
predictions.

However, at the quantum level, the nonlinear interactions appear
at the loop diagrams, so that the theory becomes strongly coupled
above the energy scale $\Lambda=(m^{4}M_{Pl})^{1/5}$, where $m$ is
the graviton mass and $M_{Pl}$ is the Planck mass
\cite{arkanihamed2003}, \cite{aubert2004}. For masses $m\sim
H_{0}$, where $H_{0}$ is the present-day value of the Hubble
parameter, the energy scale $\Lambda$ is too small, well below the
expected value, $\Lambda=(mM_{Pl})^{1/2}$. In brane-world models
(\cite{charmousis2000}, \cite{gregory2000}, \cite{kogan2000},
\cite{dvali2000}) a similar problem occurs: either they have
ghosts \cite{luty2003}, \cite{dubovsky2003}, \cite{chacko2004} and
\cite{pilo2000}, or are strong coupled at low energies
\cite{luty2003}, \cite{dubovsky2003}, \cite{chacko2004} and
\cite{rubakov2003}.

A great step forward was taken in the works \cite{arkanihamed2004}
and \cite{rubakov2004}. In reference \cite{arkanihamed2004} the
authors proposed a consistent modification of gravity in the
infrared as an analog of the Higgs mechanism in GR. In this model,
Lorentz invariance is spontaneously broken and the graviton, as a
result, acquires a mass. In reference \cite{rubakov2004} the
author introduces a Lorentz-violating massive gravity model in
which the vDVZ discontinuity, ghosts and the low strong coupling
scale are absent. In reference \cite{dubovsky2004} the author
studies the most general Lorentz-violating gravitational theory
with massive gravitons, showing that there is a number of
different regions in the mass parameter space of this theory in
which it can be described by a consistent low-energy effective
theory without instabilities and the vDVZ discontinuity.

Therefore, the theory of \emph{Massive Gravity}, as developed in
\cite{rubakov2004} and \cite{dubovsky2004}, gives rise to physical
propagating modes, and is free of the pathologies mentioned above.
It is a potential candidate to provide the proper answers to the
open questions in cosmology as mentioned earlier. There is a
number of works studying cosmology in the context of Massive
Gravity \cite{dubovsky2005a}, \cite{dubovsky2005b},
\cite{bebronne2007}; in the present work, we aim at extending this
discussion by analyzing the anisotropies of the CMB induced by
tensor perturbations in Massive Gravity. We have analyzed tensor
and vector perturbations in theories of gravitation with massive
gravitons in a previous work \cite{bessada2009}, and in this paper
we focus specifically on the signatures of the massive tensor
modes. Henceforth we shall use the terms graviton and tensor modes
interchangeably.

We know from a number of sources \cite{polnarev1985},
\cite{kamionkowski1997}, \cite{zaldarriaga1997},
\cite{baskaran2006}, that primordial GW might leave a
\emph{signature} in the anisotropies and polarization spectrum of
CMB, generated by the influence of such GWs on the photon
redshifts. So, if the gravitons do have a mass, we expect that
they will leave a different signature on the CMB anisotropy
spectrum. Therefore, the main goals of this present work are
twofold: develop solutions for massive GWs and analyze their
signatures on the CMB polarization spectrum.

To this end, the present paper is organized as follows: in section
\ref{sec:two} we review the basics of Massive Gravity and its
cosmological tensor perturbations. In section \ref{sec:three} we
start reviewing the solutions for primordial GW in GR, and right
after it we present our results in Massive Gravity; furthermore,
we compare the results of both theories for different graviton
masses, and discuss the differences between them. In section
\ref{sec:four} we review the basics of radiative transfer in the
presence of weak gravitational fields, deriving the relevant
Boltzmann equations. In section \ref{sec:five} we derive the
expressions for CMB anisotropies and polarization, with the
corresponding correlation functions. In section \ref{sec:six} we
discuss the solutions to the Volterra integral equation which
provides the functions that enables us to evaluate the
coefficients to the harmonic expansion the modes introduced in
section \ref{sec:five}. In section \ref{sec:seven} we apply all
these theoretical tools to Massive Gravity, obtaining the
individual power spectra for different wavenumbers and graviton
masses, and we compare these results with the predictions of GR.
At the end of this paper we discuss the obtained results and make
the corresponding conclusions.

\section{\label{sec:two} A quick overview of Massive Gravity}

As we have pointed out in the Introduction, the key ingredient to
construct a physically-consistent theory of gravitation with
massive gravitons lies on the spontaneous violation of the Lorentz
symmetry. As in the Higgs analog in the Standard Model of
electroweak interactions, we introduce, following
\cite{dubovsky2004} and \cite{dubovsky2005a}, a set of four scalar
Goldstone fields $\phi^{0}(x)$, $\phi^{i}(x)$, such that the
action for Massive Gravity is written as
\begin{equation}\label{massgravaction}
S=\int d^{4}x
\sqrt{-g}\left[-M_{Pl}^{2}R+\Lambda^{4}F(X,V^{i},W^{ij})+{\cal{L}}_{matter}\right],
\end{equation}
where the first term on the right-hand side represents the usual Einstein-Hilbert action, and $F$ is an arbitrary function of the metric components, their derivatives, and the Goldstone fields. The lagrangian for ordinary matter, ${\cal{L}}_{matter}$, is assumed to be minimally coupled to the metric. The simplest way to combine the derivatives of the Goldstone fields to enter the argument of $F$ is given by the set of scalar quantities 
\begin{equation}\label{xfield}
X=\Lambda^{-4}g^{\alpha\beta}\partial_{\alpha}\phi^{0}\partial_{\beta}\phi^{0},\nonumber \\
\end{equation}
\begin{equation}\label{vfield}
V^{i}=\Lambda^{-4}g^{\alpha\beta}\partial_{\alpha}\phi^{0}\partial_{\beta}\phi^{i}, \nonumber \\
\end{equation}
\begin{equation}\label{wfield}
W^{ij}=\Lambda^{-4}g^{\alpha\beta}\partial_{\alpha}\phi^{i}\partial_{\beta}\phi^{j}-\frac{V^{i}V^{j}}{X},
\end{equation}
where $\Lambda$ is the parameter which characterizes the cutoff
scale of the theory. The second term on the right-hand side of
(\ref{massgravaction}) is invariant under the spatial
reparametrization symmetry $x^{i}(t)\rightarrow
x^{i}(t)+\xi^{i}(t)$ and rotations.

We now introduce the ``vacuum" solutions for the model
(\ref{massgravaction}),
\begin{equation}\label{goldsvac}
g_{\alpha\beta}=a^{2}\eta_{\alpha\beta}, ~~~
\phi^{0}=\Lambda^{2}t, ~~~\phi^{i}=\Lambda^{2}x^{i},
\end{equation}
which corresponds to the flat FRW space; in the ``unitary gauge"
described by (\ref{goldsvac}) the action will depend solely on the
metric components. Now, in order to study linear cosmological
perturbations around a flat Friedmann-Robertson-Walker (FRW)
space, we spontaneously break the Lorentz symmetry of the model by
fixing the Goldstone fields to the vacuum (\ref{goldsvac}), so
that the only remaining perturbations are given by
\begin{equation}\label{pertmetric}
g_{\alpha\beta}=a^{2}\eta_{\alpha\beta}+\delta g_{\alpha\beta},
\end{equation}
where $\eta_{\alpha\beta}=diag\{+,-,-,-\}$, $a(\eta)$ is the scale
factor, and $\delta g_{\alpha\beta}$ is a metric perturbation
whose components are given by \cite{brandenberger1992},
\begin{equation}
\delta g_{00}=2a^{2}\varphi, ~~~ \delta g_{0i}=a^{2}(S_{i}-\partial_{i}B),\nonumber \\
\end{equation}
\begin{equation}\label{pertmetriccomp}
\delta
g_{ij}=a^{2}\left[-h_{ij}-\partial_{i}Q_{j}-\partial_{j}Q_{i}+2(\psi\delta_{ij}-\partial_{i}\partial_{j}E)\right],
\end{equation}
where $\varphi,\psi,B,E$ are scalar fields, $Q_{i}$ and $S_{i}$
are vector fields, and $h_{ij}$ is a tensor field. The constraints
satisfied by the vector and tensor fields are \cite{rubakov2004},
\cite{brandenberger1992},
\begin{equation}\label{pertmetricconstr}
{h_{ij}}^{,~j}=0,~~~ {h^{i}}_{i}=0,~~~{Q^{i}}_{,i}={S^{i}}_{,i}=0.
\end{equation}

Now, in the unitary gauge (\ref{goldsvac}) we expand
$\sqrt{-g+\delta g}$, $X(g+\delta g)$, $V^{i}(g+\delta g)$,
$W^{ij}(g+\delta g)$ and $F(g+\delta g)$ in powers of the metric
perturbation $\delta g$, and substitute these results into the
massive term in (\ref{massgravaction}), so that the lagrangian for
the second-order perturbations reads
\begin{equation}\label{masslagr1}
{\cal{L}}_{m}=\frac{M_{Pl}^{2}}{2}\left[m_{0}^{2}\delta g_{00}^{2}
+ 2m_{1}^{2}\delta g_{0i}^{2}-m_{2}^{2}\delta g_{ij}^{2}
+m_{3}^{2}\delta g_{ii}\delta g_{jj}-2m_{4}^{2}\delta g_{00}\delta
g_{ii}\right],
\end{equation}
where $m_{0}$, $m_{1}$, $m_{2}$, $m_{3}$ and $m_{4}$ are
parameters related to the function $F$ and its derivatives,
\begin{equation}
m_{0}^{2}=\frac{\Lambda^{4}}{M_{Pl}^{2}}\left[XF_{X}+2X^{2}F_{XX}\right],~~~
m_{1}^{2}=\frac{2\Lambda^{4}}{M_{Pl}^{2}}\left[-XF_{X}-WF_{W}+\frac{1}{2}XWF_{VV}\right],\nonumber
\end{equation}
\begin{equation}
m_{2}^{2}=\frac{2\Lambda^{4}}{M_{Pl}^{2}}\left[WF_{W}-2W^{2}F_{WW2}\right],~~~
m_{3}^{2}=\frac{\Lambda^{4}}{M_{Pl}^{2}}\left[WF_{W}+2W^{2}F_{WW1}\right],\nonumber \\
\end{equation}
\begin{equation}\label{massparameters}
m_{4}^{2}=-\frac{\Lambda^{4}}{M_{Pl}^{2}}\left[XF_{X}+2XWF_{XW}\right],
\end{equation}
where $W=-1/3\delta_{ij}W^{ij}$ and
\begin{equation}
F_{X}=\frac{\partial F}{\partial X},~~~F_{XX}=\frac{\partial^{2} F}{\partial X^{2}},~~~F_{VV}\delta_{ij}=\frac{\partial^{2} F}{\partial V^{i}\partial V^{j}},\nonumber \\
\end{equation}
\begin{equation}
F_{W}\delta_{ij}=\frac{\partial F}{\partial W^{ij}},~~~F_{XW}\delta_{ij}=\frac{\partial^{2} F}{\partial X\partial W^{ij}},\nonumber \\
\end{equation}
\begin{equation}\label{derivatives}
\frac{\partial^{2} F}{\partial W^{ij}\partial
W^{kl}}=F_{WW1}\delta_{ij}\delta_{kl}+F_{WW2}(\delta_{ik}\delta_{jl}+\delta_{il}\delta_{jk}).
\end{equation}
(see Appendix A in references \cite{dubovsky2005a} and
\cite{bebronne2007} for details). The spatial indices in
(\ref{masslagr1}) are summed over and, as argued in the reference
\cite{rubakov2004}, the mass parameters $m_{i}$ are proportional
to some scale denoted by $m$.

The Einstein equations for the model (\ref{massgravaction}), with
the Goldstone fields in the unitary gauge (\ref{goldsvac}), and
metric (\ref{pertmetric}) read (for computational details, see
appendix A of the references \cite{dubovsky2005a} and
\cite{bebronne2007}),
\begin{equation}
3{{\cal{H}}^{2}}=\frac{a^{2}}{M_{Pl}^{2}}(\rho_{m}+\rho_{\phi}+\rho_{\Lambda}),\nonumber \\
\label{massfriedmann1}
\end{equation}
\begin{equation}
2{{\cal{H}}'}+{{\cal{H}}^{2}}=-\frac{a^{2}}{M_{Pl}^{2}}(p_{m}+p_{\phi}+p_{\Lambda}),\nonumber \\
\end{equation}
\begin{equation}\label{massfriedmann2}
\partial_{0}(a^{3}F_{X}X^{1/2})=0,
\end{equation}
where ${\cal{H}}=a'/a$, $\rho_{m}$ and $p_{m}$ stand for the
density and pressure for the ordinary matter respectively, and
\begin{equation}\label{rhophi}
\rho_{\phi}=\Lambda^{4}XF_{X}, ~~~ p_{\phi}=\Lambda^{4}WF_{W},
\end{equation}
\begin{equation}\label{rholambda}
\rho_{\Lambda}=-\frac{\Lambda^{4}}{2}F, ~~~
p_{\Lambda}=\frac{\Lambda^{4}}{2}F.
\end{equation}
The prime represents a derivative with respect to the conformal
time $\eta$.

Once we have established the dynamical equations for the
background, let us now turn our attention to the metric
perturbations (\ref{pertmetriccomp}). The steps toward obtaining
the dynamical equations for the massive metric perturbations are
quite similar to those referred in \cite{brandenberger1992}, and
they can be found in details in the Appendix A 3 in
\cite{bebronne2007}; here we simply quote the results. Since in
this paper we are interested solely in the tensor perturbations
represented by the element $h_{ij}$ in (\ref{pertmetriccomp}), we
write its dynamical equation as \cite{bebronne2007}
\begin{equation}\label{tensorpert}
h_{ij}''-\nabla^{2}h_{ij}+2{\cal{H}}h_{ij}'+a^{2}m^{2}_{2}h_{ij}=0.
\end{equation}
Since we deal only with  the mass $m_{2}$ throughout this paper,
we henceforth drop the subscript 2 and write it simply as $m$.

To end this section let us discuss an important aspect concerning
the mass parameters of Massive Gravity. As we have pointed out in
the Introduction, there are regions in the mass parameter space in
which this theory is free of ghosts and instabilities; this means
that the mass parameters $m_{0}$, $m_{1}$, $m_{2}$, $m_{3}$ and
$m_{4}$ cannot be chosen arbitrarily, but they have to satisfy
some constraints \cite{rubakov2004}, \cite{dubovsky2004}. Since in
this paper we deal only with the mass parameter $m_{2}$, there is
a number of choices on these parameters in which the model is
physically healthy; therefore, any of these choices would produce
a physically acceptable theory. We simply assume that the mass
parameters in our work are within the region in which the
pathologies are absent.

Specific restrictions on the function $F$ are discussed in
\cite{dubovsky2005a}. In this reference, the authors demonstrate
the existence of a wide class of functions $F$ for which expanding
cosmological solutions are compatible with constant graviton
masses and allow for the effective field theory description.
Therefore, we may simply restrict $F$ in such a way the mass
$m_{2}$ is constant along the story of the universe, which we
assume to hold throughout this paper.

\section{\label{sec:three}Primordial Gravitational Waves in GR and Massive Gravity}

\subsection{\label{primgwgr}Primordial Gravitational Waves in GR}

In GR, a primordial GW is described by the cosmological tensor
perturbation whose dynamical evolution is governed by the equation
\cite{brandenberger1992}
\begin{equation}\label{tensorpertrg}
h_{ij}''-\nabla^{2}h_{ij}+2{\cal{H}}h_{ij}'=0.
\end{equation}
Before going into the Fourier space to solve this equation, it is
convenient to introduce a new parametrization into this model
\cite{bose2002}. Let us write down the present-day scale factor
$a(\eta_{0})$ as a quantity with dimension of length; then,
setting $R_{H}=c/H_{0}$ as the Hubble radius, we then define
$a(\eta_{0})=2R_{H}$. Now, since the GW wavenumber $\mathbf{k}$ is
very small for primordial GWs (in the frequency range which could
produce a signature on CMB), with wavelength comparable to the
present-day Hubble radius $R_{H}$, we introduce a dimensionless
time-independent vector $\mathbf{n}$ which has the same direction
of $\mathbf{k}$, and whose modulus is exactly the proportionality
factor between the modulus of $\mathbf{k}$ and $R_{H}$:
\begin{equation}\label{definitionofn}
n=2R_{H}k.
\end{equation}
Since the early cosmological perturbations are of
quantum-mechanical origin, we construct the tensor $h_{ij}$ as a
quantum-mechanical operator, whose Fourier expansion is given by
\begin{equation}\label{fourierexph}
h_{ij}\left(\eta,\mathbf{x}\right)=\frac{\sqrt{16\pi}\ell_{Pl}}{(2\pi)^{3/2}}\int^{\infty}_{-\infty}
\frac{d^{3}\mathbf{n}}{\sqrt{2n}}\sum_{r=1,2}[\varepsilon^{r}_{ij}(\mathbf{n})h^{r}_{n}(\eta)e^{i\mathbf{n}\cdot
\mathbf{x}}\hat{a}^{r}_{\mathbf{n}}+\varepsilon^{r\ast}_{ij}(\mathbf{n})h^{r\ast}_{n}(\eta)e^{-i\mathbf{n}\cdot
\mathbf{x}}\hat{a}^{r\dag}_{\mathbf{n}}],
\end{equation}
where $r$ stands for the polarization mode of the GW,
$\varepsilon^{r}_{ij}$ is the GW polarization tensor, and
$\ell_{Pl}$ is the Planck length. The annihilation and creation
operators $\hat{a}^{r}_{\mathbf{n}}$ and
$\hat{a}^{r\dag}_{\mathbf{n}}$ satisfy the well known commutation
relations
\begin{equation}\label{vacuumrg}
\left[\hat{a}^{r}_{\mathbf{n}},\hat{a}^{s\dag}_{\mathbf{n}'}\right]=\delta_{rs}\delta^{(3)}
(\mathbf{n}-\mathbf{n}'),
\end{equation}
and, for the vacuum state $|0\rangle$,
\begin{equation}\label{vacuumrg}
\hat{a}^{r}_{\mathbf{n}}|0\rangle=0.
\end{equation}
Now, substituting the expression (\ref{fourierexph}) into
(\ref{tensorpertrg}), and redefining the conformal time derivative
as $d/d\eta =(a/c)d/dt$, we get the equation governing the
dynamics of GR tensor modes (dropping the polarization index for a
while)
\begin{equation}\label{eqrggwamplit}
h_{n}''+2{\cal{H}}h_{n}'+n^{2}h_{n}=0.
\end{equation}
Then, defining the quantity $\mu_{n}(\eta)=a(\eta)h_{n}(\eta)$
\cite{grishchuk1974}, we obtain an equation for a parametrically
disturbed oscillator
\begin{equation}\label{eqrggwparoscill}
\mu_{n}''+\left[n^{2}-\frac{a''}{a}\right]\mu_{n}=0.
\end{equation}
To solve the simplified equation (\ref{eqrggwparoscill}) we have
to specify the scale factor. Since we are mostly interested in the
time of recombination, we use the scale factor for a flat universe
filled with radiation and matter, whose expression is
\cite{baskaran2006}
\begin{equation}\label{scalefactorrm}
a(\eta)=2R_{H}\left(\frac{1+z_{eq}}{2+z_{eq}}\right)\eta\left(\eta+\frac{2\sqrt{2+z_{eq}}}{1+z_{eq}}\right),
\end{equation}
where $z_{eq}$ is the redshift associated with the epoch of
radiation-matter equality, whose value is $z_{eq}\sim 3\times
10^{3}$, and the corresponding conformal instant $\eta_{eq}$ is
given by
\begin{equation}
\eta_{eq}=(\sqrt{2}-1)\frac{\sqrt{2+z_{eq}}}{1+z_{eq}}\sim 7.6
\times 10^{-3}.
\end{equation}

It is easy to see that (\ref{scalefactorrm}) reduces to a scale
factor for a radiation-dominated and matter-dominated universe
\begin{eqnarray}\label{limitsofa}
a(\eta) &=& \frac{4R_H}{\sqrt{1+z_{eq}}} \eta,
~~\eta\leq\eta_{eq};\nonumber \\a(\eta) &=& 2R_H \left(\eta
+\eta_{eq}\right)^{2},~~\eta\geq\eta_{eq}
\end{eqnarray}
respectively, so that it comprises the whole period we are
interested in.

Substituting the scale factor (\ref{scalefactorrm}) into
(\ref{eqrggwparoscill}), we can obtain exact analytical solutions
for the functions $\mu_{n}(\eta)$ \cite{bose2002}. Following
\cite{baskaran2006}, we normalize the GW amplitudes $h_{n}(\eta)$
in terms of its value at $\eta_{r}=10^{-6}$ (in terms of redshift,
$z_{r}\sim 3\times 10^{7}$); the resulting numerical solutions are
displayed in the figure \ref{fig1}.
\smallskip
\FIGURE{\epsfig{file=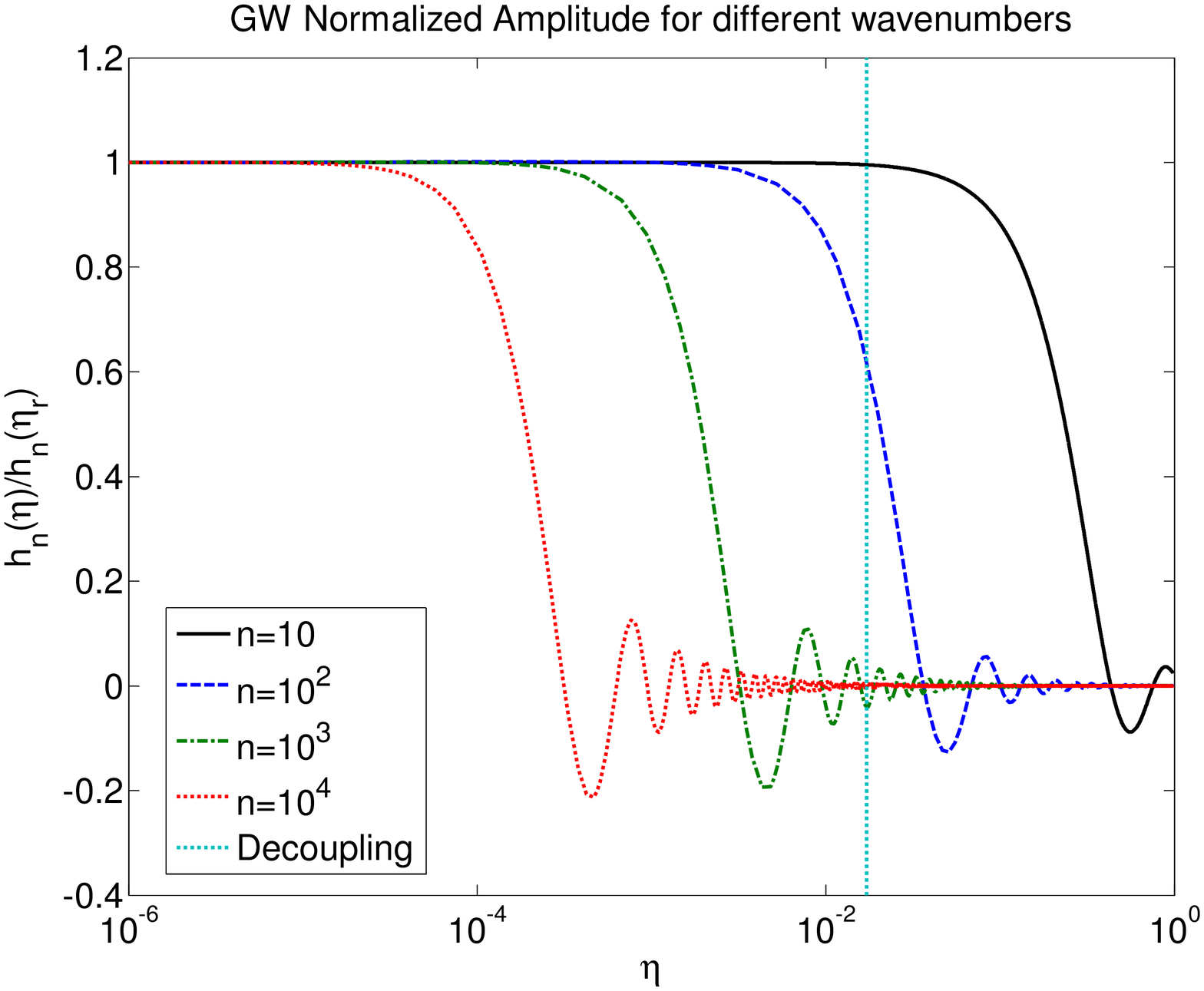,width=9cm}
        \caption[]{The time evolution of the
normalized GW amplitudes $h_{n}(\eta)/h_{n}(\eta_{r})$. Compare
with Figure 1 of \cite{baskaran2006}.}%
    \label{fig1}}

\subsection{\label{primgwmassive}Primordial Gravitational Waves in Massive Gravity}

Once we have reviewed the properties and evolution of GW
amplitudes in GR, let us now analyze the same issues in Massive
Gravity. First of all, we treat the tensor perturbations in the
massive case quantum-mechanically, so that the Fourier expansion
for the massive tensor field $h_{ij}\left(\eta,\mathbf{r}\right)$
has an analog expression as in (\ref{fourierexph}):
\begin{eqnarray}\label{fourierexphmass}
h^{(m)}_{ij}\left(\eta,\mathbf{x}\right)&=&\frac{\sqrt{16\pi}\ell_{Pl}}{(2\pi)^{3/2}}\int^{\infty}_{-\infty}
\frac{d^{3}\mathbf{n}}{\sqrt{2E_{\mathbf{n}}}}\sum_{r=1,2}[\varepsilon^{(m)r}_{ij}(\mathbf{n})h^{(m)r}_{n}(\eta)e^{i\mathbf{n}\cdot
\mathbf{x}}\hat{a}^{(m)r}_{\mathbf{n}}\nonumber
\\&+&\varepsilon^{(m)r\ast}_{ij}(\mathbf{n})h^{(m)r\ast}_{n}(\eta)e^{-i\mathbf{n}\cdot
\mathbf{x}}\hat{a}^{(m)r\dag}_{\mathbf{n}}],
\end{eqnarray}
where the superscript $(m)$ stands for massive, and
$E_{\mathbf{n}}$ denotes the energy of the mode $\mathbf{n}$. Now,
plugging (\ref{fourierexphmass}) into (\ref{tensorpert}) we get
\begin{equation}\label{eqrggwamplitmassive}
h^{(m)''}_{n}+2{\cal{H}}h^{(m)'}_{n}+\left(n^{2}+m^{2}a^{2}\right)h^{(m)}_{n}=0.
\end{equation}
In (\ref{eqrggwamplitmassive}) we have dropped the GW polarization
indices $r$ because we are going to treat only the tensor modes in
this paper. Using the same strategy as in the GR case, we
introduce a function
$\mu^{(m)}_{n}(\eta)=a(\eta)h^{(m)}_{n}(\eta)$, so that equation
(\ref{eqrggwamplitmassive}) becomes
\begin{equation}\label{eqrggwparoscillmassive}
\mu^{(m)''}_{n}+\left[n^{2}+m^{2}a^{2}-\frac{a''}{a}\right]\mu^{(m)}_{n}=0.
\end{equation}

As we have discussed in the section above, the scale factor
(\ref{scalefactorrm}) represents very well the periods of the
universe considered in this paper, so that it makes sense to
employ it in Massive Gravity as well, since we may expect that the
contribution of massive gravitons to the expansion of the universe
is negligible in its early epochs; then, as a first approximation,
we may neglect the contribution of the components $\rho_{\phi}$,
equation (\ref{rhophi}), and $\rho_{\Lambda}$, equation
(\ref{rholambda}), in (\ref{massfriedmann1}).

Now, using the above arguments and consequently the scale factor
(\ref{scalefactorrm}), we can solve numerically equation
(\ref{eqrggwparoscillmassive}) for different wavenumbers $n$ and
masses $m$. We choose the graviton masses $m$ using the following
argument: in GR, only GW with frequencies $\nu$ within the range
$10^{-15}Hz$ to $10^{-18}Hz$ may leave a signature on CMB
polarization \cite{kamionkowski1998}; these frequencies correspond
to wavenumbers $k$ within the range $10^{-25}cm^{-1}$ ($n\sim
5\times 10^{3}$) to $10^{-28}cm^{-1}$ ($n\sim 10$). For Massive
Gravity, we use the same values for $k$, but now we vary the
frequencies in order to obtain constant nonzero graviton masses
through the dispersion relation
\begin{equation}\label{disprelation}
\omega^{2}=k^{2}+m^{2},
\end{equation}
which comes straight from (\ref{tensorpert} ), where now
$\omega=2\pi \nu$. As a result, we find that if the values of the
mass $m$ lie within the range $10^{-66}$ - $10^{-62}g$, the
corresponding frequencies have values very close to the expected
in GR. In particular, we've found that if the graviton mass is
$m=10^{-66}g\sim 10^{-29}cm^{-1}$, the behavior of the GWs in
Massive Gravity is exactly \emph{the same} of GWs in GR.
Therefore, if the graviton mass is equal or less than the graviton
mass limit $m_l=10^{-66}g$, \emph{the effects of Massive Gravity
are indistinguishable from GR}.

It is important to mention that there has been a lot of efforts to constrain the masses of the tensor modes over the past few decades.
For instance, Goldhaber and Nieto
\cite{goldhaber1974} have found a limit $m < 2.0 \times 10^{-62}g$
analyzing the motion of galaxies in clusters. Later on, Talmadge
\emph{et al.} \cite{talmadge1988} studied the variations of
Kepler's third law when compared with the orbits of Earth and
Mars, and found a limit $m < 7.68 \times 10^{-55}g$. Recently,
Finn and Sutton \cite{Finn2002} calculated the decay of the
orbital period of the binary pulsars PSR B1913+16 (Hulse and
Taylor pulsar) and PSR B1534+12 due to emission of massive
gravitons, and found $m < 1.4\times 10^{-52} g$. Cooray and Seto \cite{Cooray:2003cv} investigated the variation of the speed of gravity when compared
to the speed of light due to a massive tensor mode, and determined an upper limit of $\sim 10^{-56}g$ for its mass, by using the measurements of
a sample of close white dwarf binaries detectable with the Laser Interferometer Space Antenna (LISA), together with a optical light curve data related
to binary eclipses from meter-class telescopes for the same sample. A recent and comprehensive review of the methods to determine the bounds for the
masses of gravitons and photons can be found in \cite{Goldhaber:2008xy}, which we refer to for further details.

Since we are interested in investigating signatures of massive
gravitons, we shall consider only graviton masses higher than the
limit $m=10^{-66}g$; the numerical solutions to the massive tensor
perturbation equations (\ref{eqrggwparoscillmassive}) are depicted
in the Figure \ref{fig2} below. For sake of comparison we depict
the general-relativistic GW amplitudes in each graph as well. We
have used the same normalization as \cite{baskaran2006}, and the
plots start at $\eta=\eta_{r}=10^{-6}$. The mass $m=2.843\times
10^{-28}cm^{-1}$ correspond to $m= 10^{-65}g$, and so forth.

Let us now analyze in detail the behavior of massive gravitons in the light of equation (\ref{eqrggwparoscillmassive}). In the very early universe,
before the time of equality radiation-matter, the value of $a(\eta)$ is very low, and then the $m^2a^2$ on the left-hand side of
(\ref{eqrggwparoscillmassive}) can be dropped; therefore, we recover the characteristic tensor mode equation of GR, (\ref{eqrggwparoscill}),
and the behavior of massless and massive gravitons are the same. On superhorizon
scales, $n\ll a''/a$, the resulting equation for the tensor modes is
\begin{equation}\label{eqsuperh}
\mu^{(m)''}_{n}-\frac{a''}{a}\mu^{(m)}_{n}=0,
\end{equation}
whose solution is given by $\mu^{(m)}_{n}=f(n)a$, which means that the tensor amplitudes are ``frozen", no matter the gravitons are massless or
not. This particular behavior can be clearly seen from figures \ref{fig2} - \ref{fig7}, where the amplitudes are constant for all the modes considered
prior to decoupling.

However, as the universe evolves, the tensor modes ``fall" into the horizon, so that their amplitudes are no longer constant; on subhorizon scales,
$n\gg a''/a$, we can neglect the effect of the term $a''/a$, so that we are left with
\begin{equation}\label{eqsubh}
\mu^{(m)''}_{n}+\left[n^{2}+m^{2}a^{2}\right]\mu^{(m)}_{n}=0.
\end{equation}
On subhorizon scales the massive term becomes dominant over low values of $n$, so that it ``enforces" the tensor modes to fall into the horizon
earlier
than in the massless case. It is clear from equation (\ref{eqsubh}) that the heavier the gravitons, the earlier their modes fall into the horizon.
\smallskip
\DOUBLEFIGURE[t]{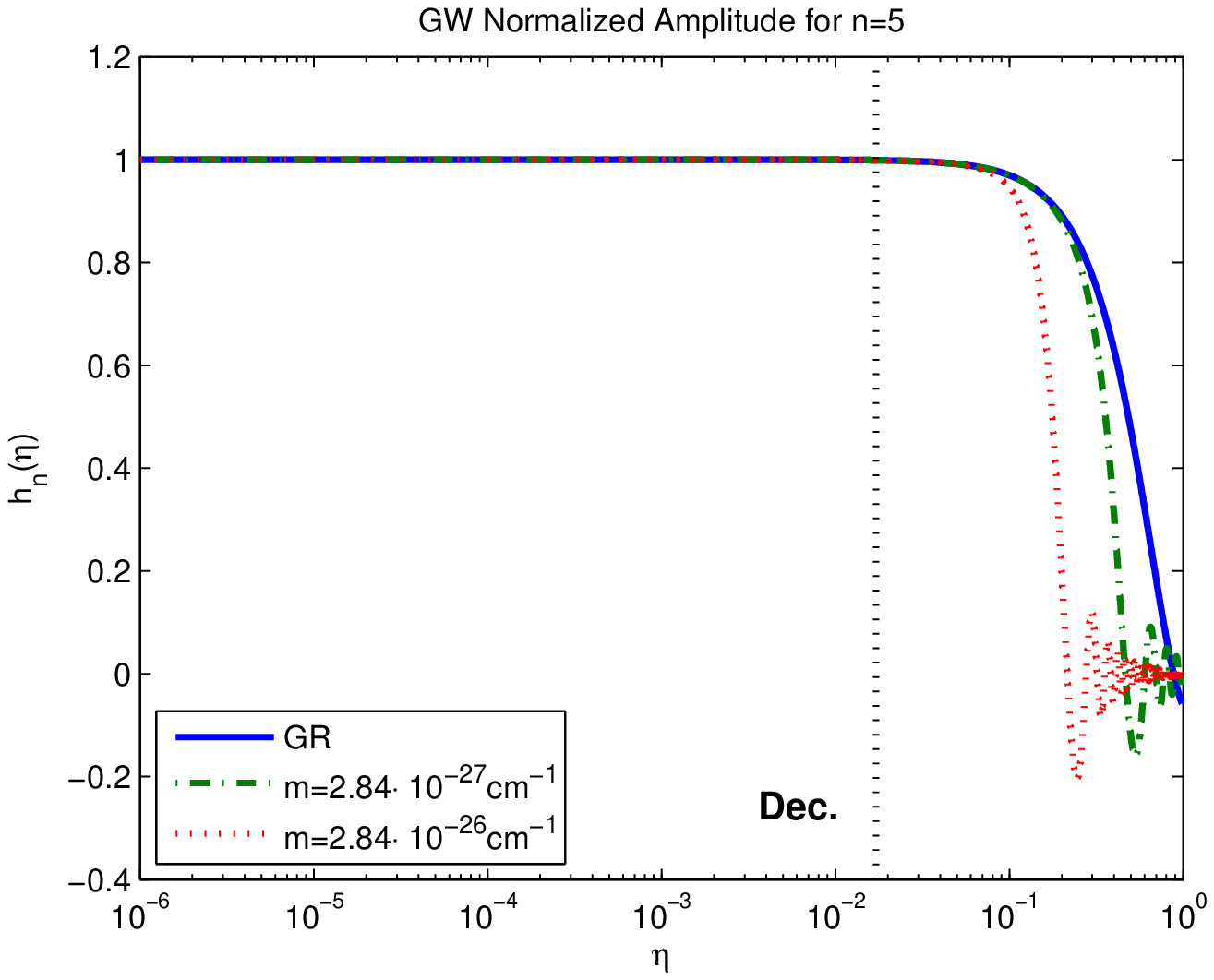, width=.53\textwidth}
{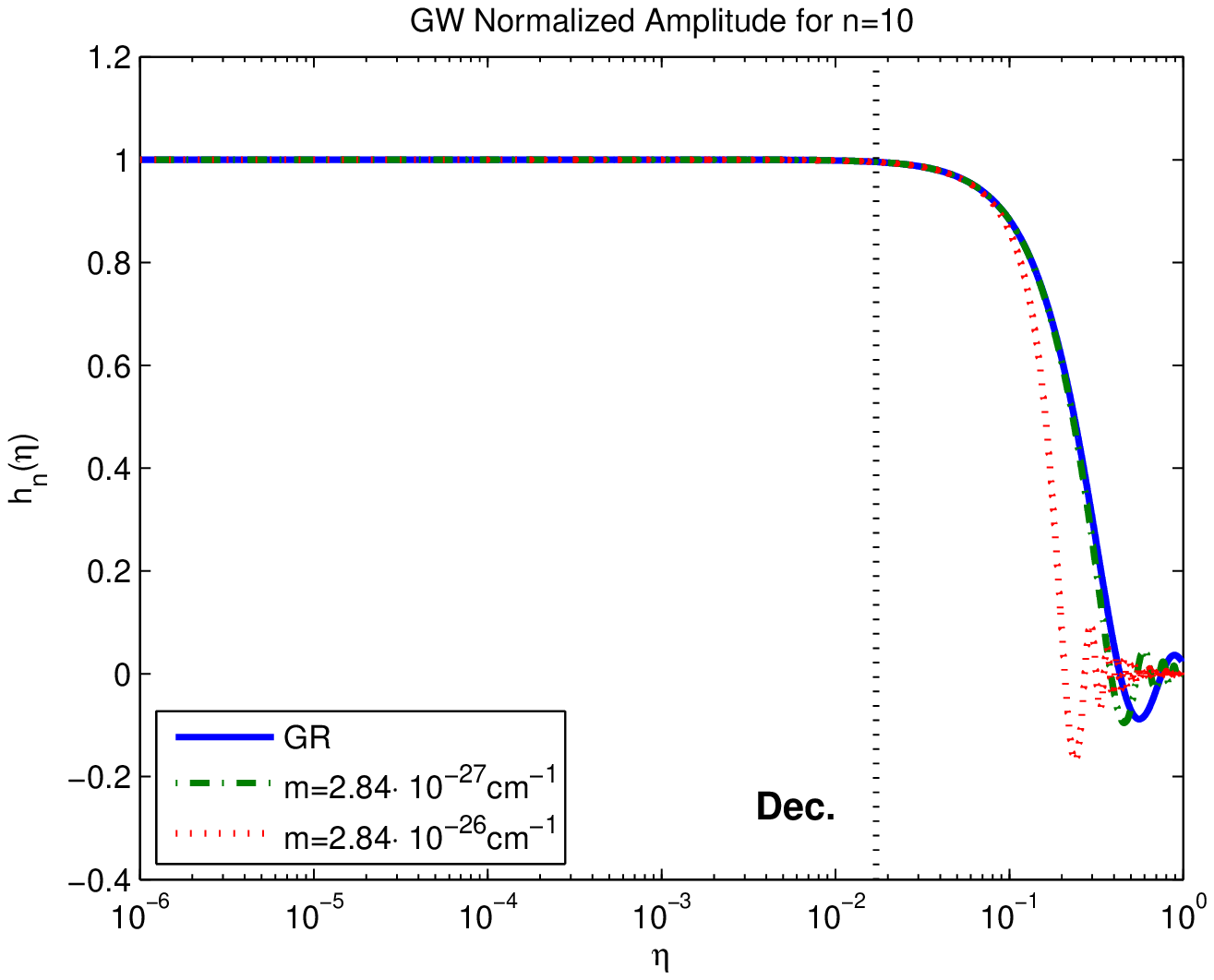, width=.53\textwidth}{The time evolution of
the normalized GW amplitudes for $n=5$.\label{fig2}}{For
$n=10$.\label{fig3}} \DOUBLEFIGURE[t]{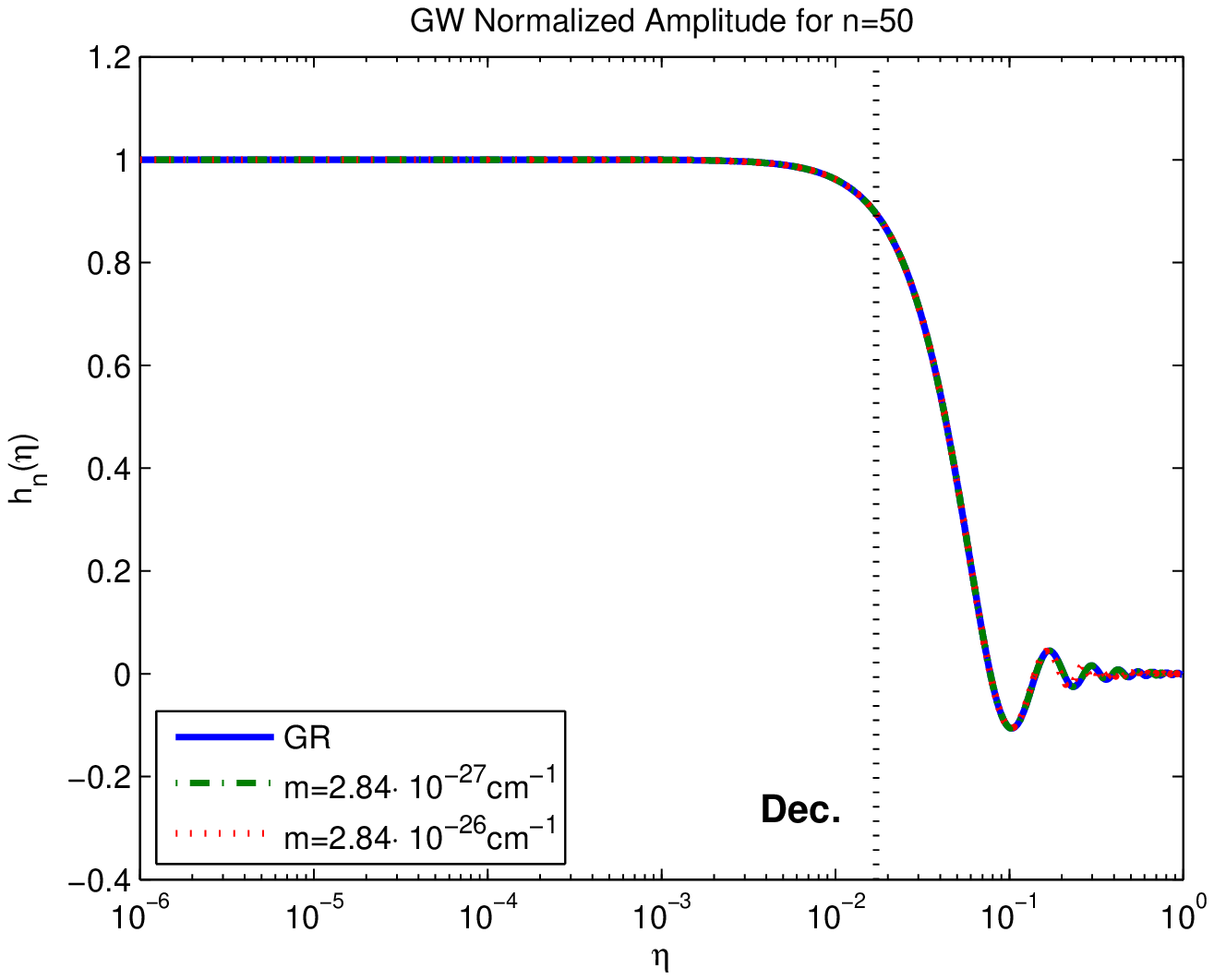,
width=.53\textwidth} {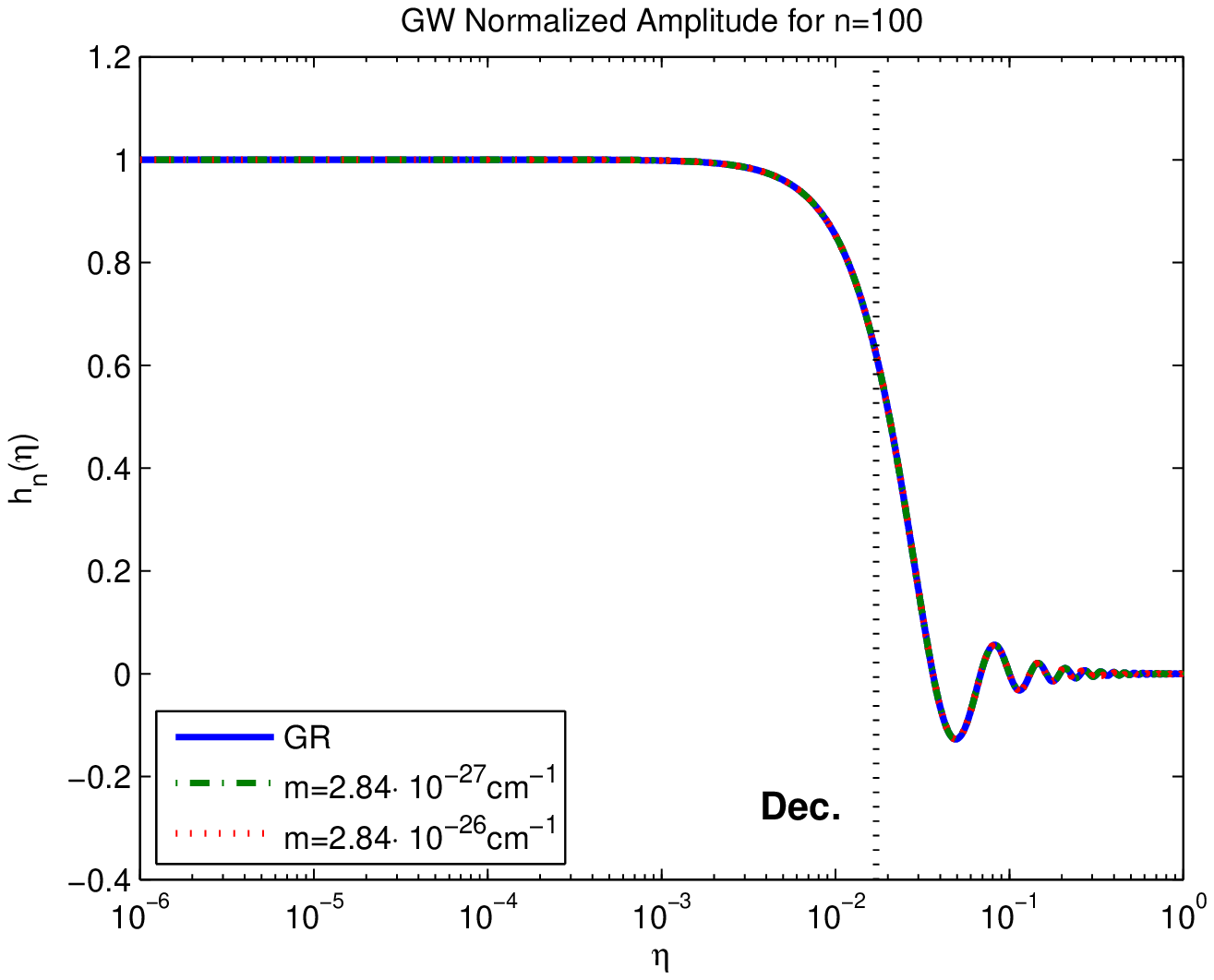, width=.53\textwidth}{For
$n=50$.\label{fig4}}{For $n=100$.\label{fig5}}
\DOUBLEFIGURE[t]{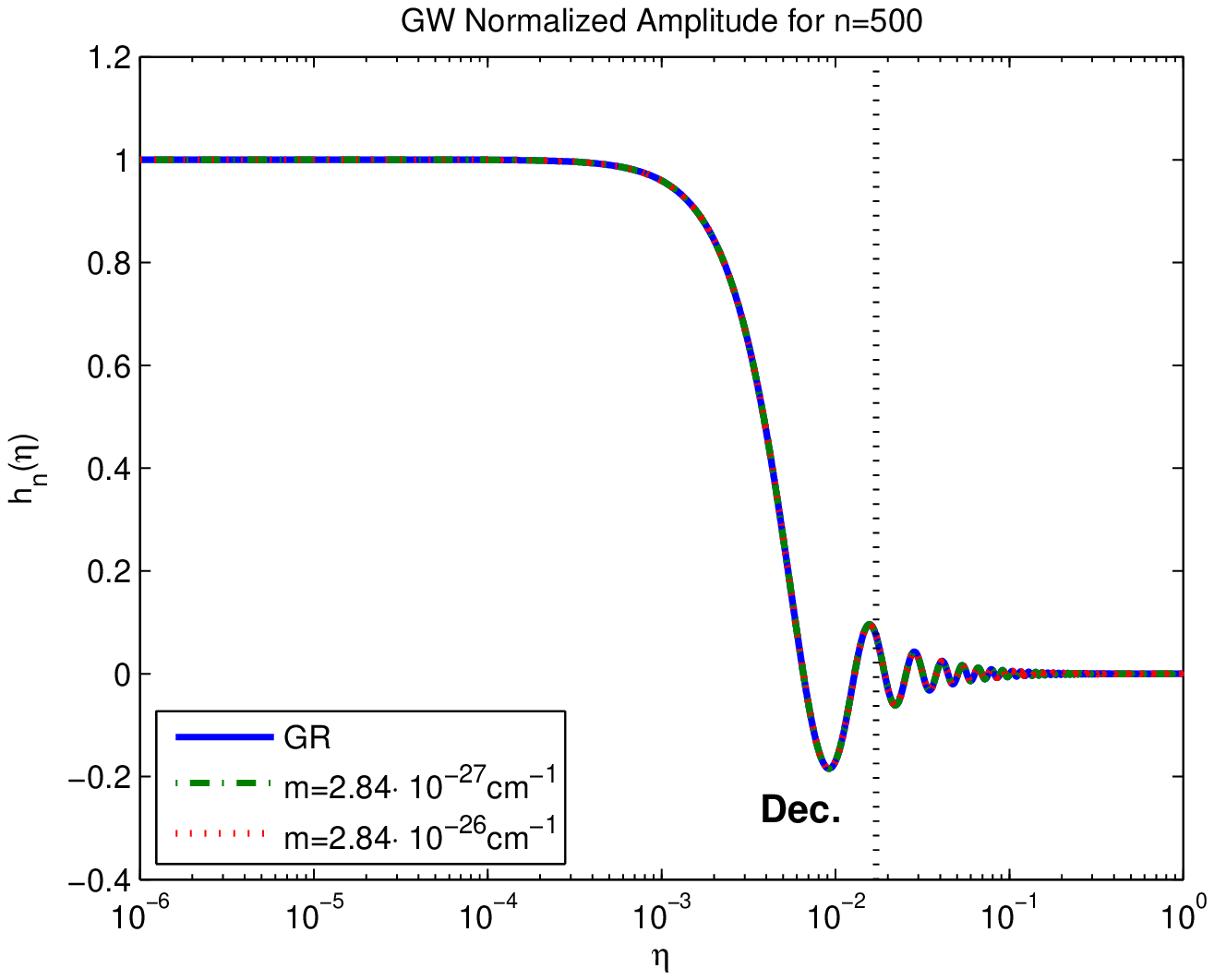, width=.53\textwidth}
{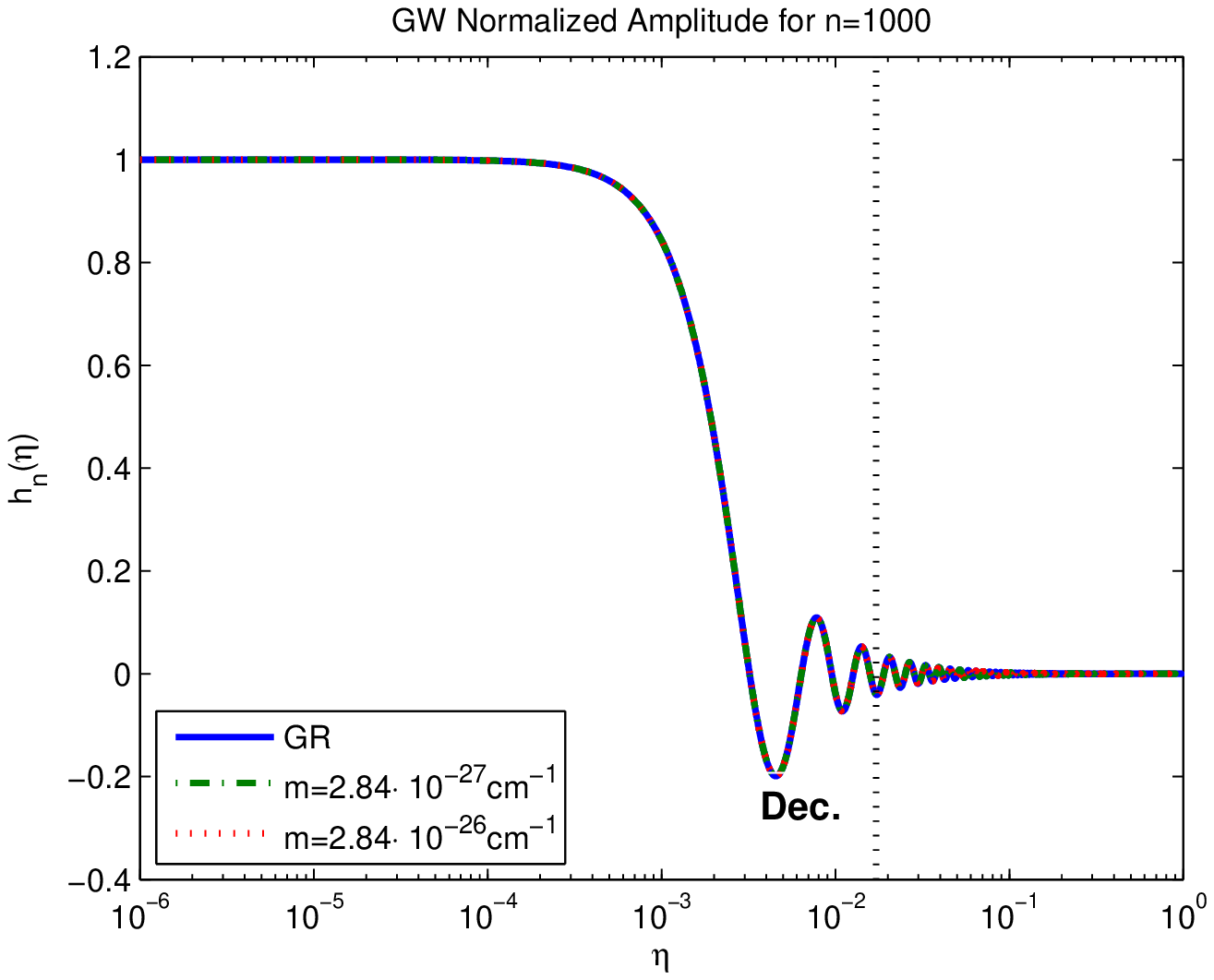, width=.53\textwidth}{For
$n=500$.\label{fig6}}{For $n=1000$.\label{fig7}}
\smallskip
\FIGURE{\epsfig{file=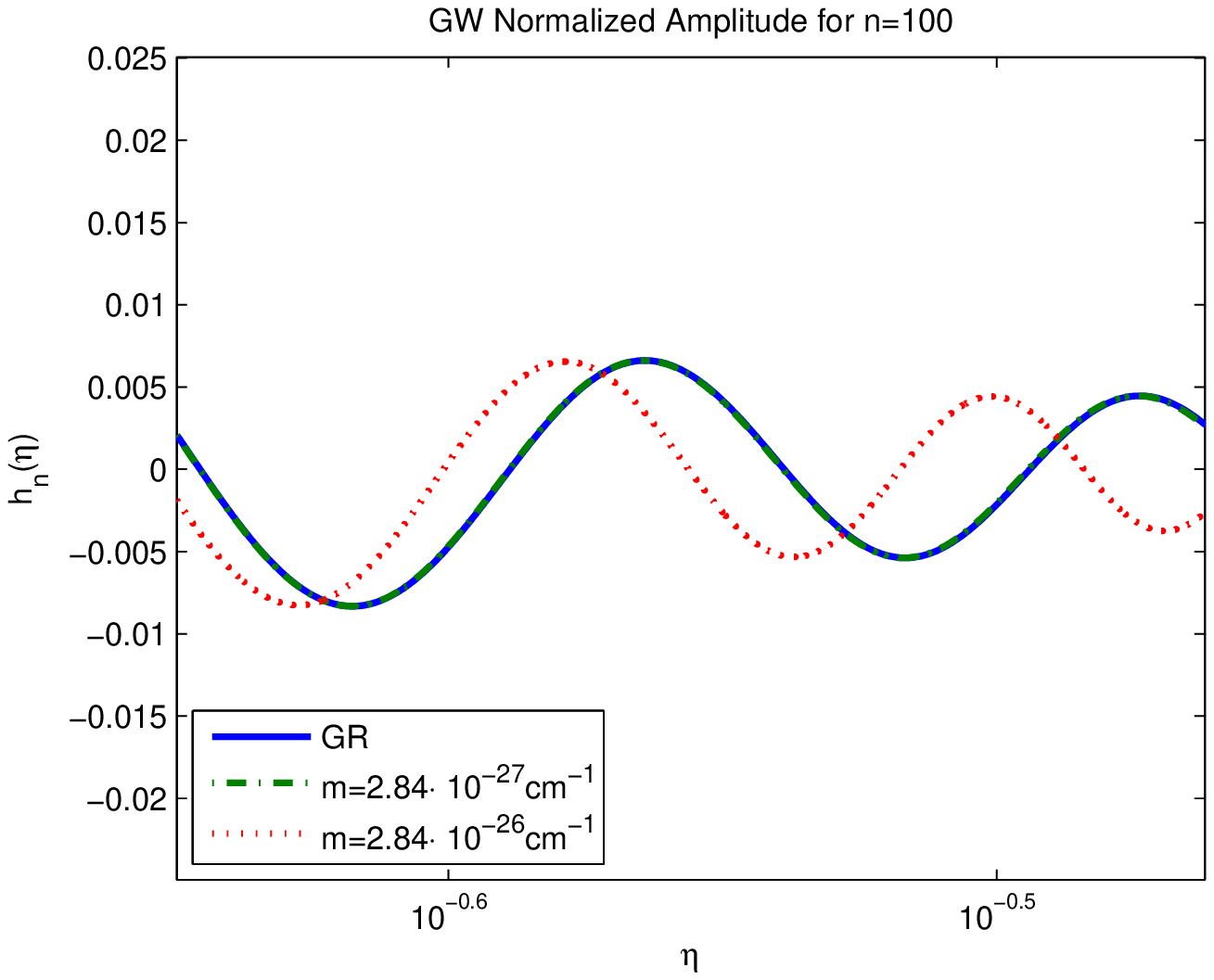,width=9cm}
        \caption[]{The ``tail of figure \ref{fig4} zoomed in, showing the phase difference in the tensor modes at very low redshifts for both massless
        and massive gravitons.}%
    \label{fig8}}

This effect can be clearly seen in the figures \ref{fig2}, \ref{fig3} and \ref{fig4}, where the $n$ values are sufficiently low to account
for this effect. However, for larger values of $n$, this effect weakens, since $n^2\gg m^2a^2$ in the time of decoupling, and the massive term will
be predominant only for low redshifts, as can be seen in figures \ref{fig5}, \ref{fig6} and \ref{fig7}.

In particular, for low $n$ (corresponding to tensor modes with long wavelengths), its constant contribution to (\ref{eqsubh}) can be completely
neglected, so that we are left with
\begin{equation}\label{eqsubhln}
\mu^{(m)''}_{n}+\left[n^{2}+m^{2}a^{2}-\frac{a''}{a}\right]\mu^{(m)}_{n}=0,
\end{equation}
and then the oscillatory behavior is strikingly different from the
massless case, as shown in figures \ref{fig2}, \ref{fig3}. For
higher $n$ (that is, tensor modes with short wavelengths), tough,
this effect is not so strong, but induces a slight phase
difference in the oscillatory behavior of the tensor modes. Such
phase difference is stronger for higher masses; as an example of
it, we have zoomed in the ``tail" of figure \ref{fig3} to show
this fact. This is presented in figure \ref{fig8}.

Hence, from this analysis we may conclude that the tensor modes of
Massive Gravity behave similarly to the massless modes of GR, but
the heavier the tensor modes are, the more distinct are their
physical evolution if compared to the massless modes.
Nevertheless, if massive gravitons do exist, they likely have left
a signature on some physical observable; then, by comparing the
predicted signatures of the massless and the massive modes with
the observed ones, one should be able to determine whether they
possess or not a nonzero mass. In our view, the best laboratory to
test this assumption is the anisotropies measurement of CMB, for
reasons that will become clear in section \ref{sec:seven}. Before
doing so, though, we take some time to review the basic aspects of
the theory of CMB anisotropies and polarization.

\section{\label{sec:four}The Radiative Transfer Equation in the
 presence of Weak Gravitational Fields - an overview}

The first account of the effect of primordial GWs in polarizing
the CMB photons was introduced in a seminal paper by Polnarev
\cite{polnarev1985}. To begin with, let us consider a given beam
of radiation characterized by its \textit{Stokes parameters}
\cite{chandrasekhar1960} $\{I,Q,U,V\}$, where $I$ is the total
intensity of the wave, the parameters $Q$ and  $U$ measure the
linear polarization of the wave, and $V$ measures its circular
polarization. They are integrated over all radiation frequencies,
so that there is a set of Stokes parameters for each monochromatic
component wave of the radiation beam
$\{I(\nu,\theta,\varphi),Q(\nu,\theta,\varphi),U(\nu,\theta,\varphi),V(\nu,\theta,\varphi)\}$.
Associated with the Stokes parameters are the components of the
photon distribution function $n$, which can be cast in a symbolic
vector of the form \cite{polnarev1985}, \cite{chandrasekhar1960},
\begin{equation}\label{stokespar}
\hat{\mathbf{n}}=\frac{1}{2}\frac{c^{2}}{h\nu^{3}}\left(
\begin{array}{c}
I+Q\\
I-Q\\
-2U\\
\end{array}
\right).
\end{equation}
Now, the transfer equation for the photon distribution functions
subject to a weak GW-field is given by
\begin{equation}\label{boltz1}
\frac{\partial \hat{\mathbf{n}}}{\partial
\eta}+{\hat{e}}^{i}\frac{\partial \hat{\mathbf{n}}}{\partial
x^{i}}+\frac{\partial \hat{\mathbf{n}}}{\partial \nu}\frac{d
\nu}{d \eta}=C\left[\hat{\mathbf{n}}\right],
\end{equation}
where ${\hat{e}}^{i}$ is the unit vector along the photon
geodesic, and $C\left[\hat{\mathbf{n}}\right]$ is the scattering
term given by
\begin{equation}\label{collterm}
C[{\mathbf{n}}]=-\sigma_{T}N_{e}a(\eta)\bigg\{\hat{\mathbf{n}}(\eta,\mathbf{r},\nu,\mu,\varphi)
-\frac{1}{4\pi}\int^{1}_{-1}d\mu'd\varphi ~
P\left(\mu,\varphi,\mu',\varphi'\right)\hat{\mathbf{n}}(\eta,\mathbf{r},\nu,\mu',\varphi')\bigg\},
\end{equation}
$P\left(\mu,\varphi,\mu',\varphi'\right)$ is the scattering matrix
given by \cite{chandrasekhar1960}
\begin{equation}\label{collterm1}
P\left(\mu,\varphi,\mu',\varphi'\right)=Q\bigg\{P^{0}\left(\mu,\mu'\right)+\sqrt{1-\mu^{2}}
\sqrt{1-\mu'^{2}}P^{1}\left(\mu,\varphi,\mu',\varphi'\right) +
P^{2}\left(\mu,\varphi,\mu',\varphi'\right)\bigg\},
\end{equation}
where
\begin{equation}\label{matscat1}
Q=\left(
\begin{array}{cccc}
1 & 0 & 0 & 0 \\
0 & 1 & 0 & 0 \\
0 & 1 & 0 & 0 \\
0 & 0 & 0 & 2 \\
\end{array}
\right),
\end{equation}
\begin{equation}\label{matscat2}
P^{0}=\frac{3}{4}\left(
\begin{array}{cccc}
2(1-\mu^{2})
(1-\mu'^{2})+\mu^{2}\mu'^{2} & \mu^{2} & 0 & 0 \\
\mu'^{2} & 1 & 0 & 0 \\
0 & 0 & 0 & 0 \\
0 & 0 & 0 & \mu \mu' \\
\end{array}
\right),
\end{equation}
\begin{equation}\label{matscat3}
P^{1}=\frac{3}{4}\left(
\begin{array}{cccc}
4\mu \mu' \cos \psi & 0 & -2\mu \sin \psi & 0 \\
0 & 0 & 0 & 0 \\
2\mu' \sin \psi & 0 & \cos \psi  & 0 \\
0 & 0 & 0 & \cos \psi \\
\end{array}
\right),
\end{equation}
\begin{equation}\label{matscat4}
P^{2}=\frac{3}{4}\left(
\begin{array}{cccc}
\mu^{2} \mu'^{2} \cos 2\psi & -\mu^{2}\cos 2\psi & -\mu^{2} \mu'
\sin
 2\psi & 0 \\
-\mu'^{2}\cos 2\psi & \cos 2\psi & \mu' \sin 2\psi & 0 \\
\mu \mu'^{2} \sin 2\psi & -\mu \sin 2\psi & \mu \mu' \cos 2\psi  & 0 \\
0 & 0 & 0 & 0 \\
\end{array}
\right),
\end{equation}
$\sigma_{T}$ is the Thomson scattering cross-section,
$N_{e}(\eta)$ is the number of free electrons in the unit comoving
volume, $\mu=\cos \theta$, and we have defined
$\psi:=\varphi-\varphi'$ \cite{polnarev1985}.

In order to get the expression of the Boltzmann equation for our
problem, let us decompose the vector ${\hat{\mathbf{n}}}$ into its
zeroth-order contribution, ${\hat{\mathbf{n}}}^{(0)}$ \emph{i.
e.}, in the absence of GW, and its first-order correction
${\hat{\mathbf{n}}}^{(1)}$,
\begin{equation}\label{decomposn}
{\hat{\mathbf{n}}}={\hat{\mathbf{n}}}^{(0)}+{\hat{\mathbf{n}}}^{(1)},
\end{equation}
where $n^{(0)}$ is the blackbody radiation function
\begin{equation}
n^{(0)}(\nu)=\frac{1}{e^{h\nu /kT_{0}}-1}, \label{blackbodyrad}
\end{equation}
and $T_{0}\sim 2.725$ K is the present-day value of the CMB
temperature. Since equation (\ref{boltz1}) is linear, we can
expand ${\hat{\mathbf{n}}}^{(1)}$ in the same way as we did in
(\ref{fourierexph}),
\begin{equation}
{\hat{\mathbf{n}}}^{(1)}(\eta,\mathbf{x},\nu,\hat{\mathbf{e}})=\frac{\sqrt{16\pi}\ell_{Pl}}{(2\pi)^{3/2}}
\int^{\infty}_{-\infty}\frac{d^{3}\mathbf{n}}{\sqrt{2n}}\sum_{r=+,\times}\bigg\{{\hat{\mathbf{n}}}^{(1)}_{\mathbf{n},r}
(\eta,\nu,\hat{\mathbf{e}})e^{i\mathbf{n}\cdot
\mathbf{x}}\hat{a}^{r}_{\mathbf{n}}
+{\hat{\mathbf{n}}}^{(1)\ast}_{\mathbf{n},r}
(\eta,\nu,\hat{\mathbf{e}})e^{-i\mathbf{n}\cdot
\mathbf{x}}\hat{a}^{r\dag}_{\mathbf{n}}\bigg\};
\label{fourierexpn}
\end{equation}
now, introducing the basis for Thomson scattering
\cite{polnarev1985},
\begin{equation}\label{basiscirc}
{\hat{a}}_{r}(\mu,\varphi)=\frac{1}{2}\left(1-\mu^{2}\right)e^{\pm
2i\varphi}~\hat{\mathbf{u}},~~~
{\hat{b}}_{r}(\mu,\varphi)=\frac{1}{2}\left(
\begin{array}{c}
\left(1+\mu^{2}\right)\\
-\left(1+\mu^{2}\right)\\
\mp 4i\mu\\
\end{array}
\right)e^{\pm 2i\varphi},
\end{equation}
where $r=1$ corresponds to a left-hand polarization, and $r=2$ to
the right-hand one, we may further expand the functions
${\hat{\mathbf{n}}}^{(1)}_{\mathbf{n},r}
(\eta,\nu,\hat{\mathbf{e}})$ as
\begin{equation}
{\hat{\mathbf{n}}}^{(1)}_{\mathbf{n},r}(\eta,\nu,\mu,\varphi)=\frac{1}{2}f(\nu)
[\alpha_{\mathbf{n},r}(\eta,\mu)\hat{a}_{r}(\mu,\varphi)+\beta_{\mathbf{n},r}(\eta,\mu)
\hat{b}_{r}(\mu,\varphi)], \label{vectorn}
\end{equation}
where $\alpha_{\mathbf{n},r}(\eta,\mu)$ and
$\beta_{\mathbf{n},r}(\eta,\mu)$ are functions to be determined by
the solutions to the Boltzmann equation, and
\begin{equation}\label{deff}
f(\nu)=\nu\frac{dn^{(0)}}{d\nu}.
\end{equation}
Now, plugging relation (\ref{vectorn}) into (\ref{boltz1}), using
the geodesic equation for the photon,
\begin{eqnarray}\label{geodgen}
\frac{d\nu}{d\lambda}&=&-\nu
\left[\mathcal{H}+\frac{1}{2}\frac{\partial h_{ij}}{\partial
\eta}p^{i}p^{j}\right]\frac{d \eta}{d\lambda},
\end{eqnarray}
where $\lambda$ is an affine parameter, and $\mathcal{H}$ is the
Hubble parameter in conformal time, and expanding the photon
momenta in spherical coordinates around the GW direction
$\hat{\mathbf{n}}$, we obtain the Boltzmann equations for the
radiative transfer in the presence of weak gravitational fields
(dropping the indices $\mathbf{n},r$ for the sake of simplicity)
\cite{polnarev1985}, \cite{baskaran2006},
\begin{equation}\label{boltzeq1}
\frac{\partial}{\partial
\eta}\beta(\eta,\mu)+[q(\eta)+in\mu]\beta(\eta,\mu)=\frac{3}{16}q(\eta)I(\eta),
\end{equation}
\begin{equation}\label{boltzeq2}
\frac{\partial}{\partial
\eta}\xi(\eta,\mu)+[q(\eta)+in\mu]\xi(\eta,\mu)=\frac{d}{d
\eta}h(\eta),
\end{equation}
where we have defined
\begin{equation}\label{defksi}
\xi(\eta,\mu)=\alpha(\eta,\mu)+\beta(\eta,\mu),
\end{equation}
and
\begin{equation}\label{defi}
{\mathcal{I}}(\eta)=\int^{1}_{-1}d\mu'\left[(1+\mu'^{2})^{2}\beta(\eta,\mu')-\frac{1}{2}(1-\mu'^{2})^{2}\xi(\eta,\mu')\right],
\end{equation}
and introduced \emph{scattering rate} $q(\eta)$ defined by
\begin{equation}\label{scattrate}
q(\eta)=\sigma_{T}N_{e}(\eta)a(\eta).
\end{equation}

The solutions to the Boltzmann equations (\ref{boltzeq1}) and
(\ref{boltzeq2}), given by the functions $\alpha(\eta,\mu)$ and
$\beta(\eta,\mu)$, are the essential elements for the computation
of the anisotropies and polarization of the CMB, as we sketch in
the next section.

\section{\label{sec:five} Harmonic analysis on a 2-sphere}

To begin with, let us first construct the polarization tensor
associated with the Stokes parameters $Q(\theta,\varphi)$ and
$U(\theta,\varphi)$, where the coordinates $(\theta,\varphi)$
describe the position of a given region of the sky. The Stokes
parameters $Q$ and $U$ can be cast into the symmetric trace-free
(STF) \emph{polarization tensor} \cite{kamionkowski1997},
\cite{kamionkowski2004}
\begin{equation}\label{spherestfpoltensor}
{\cal{P}}_{ab}(\theta,\varphi)=\frac{1}{2}\left(
\begin{array}{cc}
Q & -U\sin \theta \\
-U\sin \theta & -Q\sin^{2}\theta \\
\end{array}
\right).
\end{equation}

On the two-sphere tensor analysis can be easily implemented; the
``divergence" and ``curl" of a symmetric rank-2 tensors are
respectively given by ${T^{ab}}_{:ab}$ and
${T^{ab}}_{:ac}{\varepsilon^{c}}_{b}$, where ``$:$" denotes
covariant differentiation, $g_{ab}$ and $\varepsilon_{ab}$ are
respectively the two-dimensional metric and antisymmetric tensors
on the 2-sphere, given by
\begin{equation}\label{metriceps2sphere}
g_{ab}(\theta,\varphi)=\left(
\begin{array}{cc}
1 & 0 \\
0 & \sin^{2} \theta \\
\end{array}
\right),~~~ \varepsilon_{ab}(\theta,\varphi)=\sin \theta \left(
\begin{array}{cr}
0 & -1 \\
1 & 0 \\
\end{array}
\right).
\end{equation}
With these elements on hand, we introduce invariants which can be
built up from the polarization tensor ${\cal{P}}_{ab}$ and its
derivatives. From the symmetric tensor on the 2-sphere we
construct two of the invariants, and from the second derivatives
we construct the other two in the form of a ``divergence" and a
``curl" \cite{baskaran2006},
\begin{eqnarray}\label{defebmodes}
I&=&g^{ab}P_{ab},~~~
V=i\varepsilon^{ab}P_{ab},~~~E=-2{{\cal{P}}_{ab}}^{:ab}, ~~~
B=-2{{\cal{P}}_{ab}}^{:bc}{\varepsilon^{a}}_{c}.
\end{eqnarray}
With these invariants we get a very convenient way to completely
characterize the radiation beam, since they do not depend on the
reference frame chosen. We now proceed to expand the invariants
$(I,E,B,V)$ in spherical harmonics in order to perform an analysis
on the each multipole of the radiation field \cite{baskaran2006}
\begin{eqnarray}
I(\theta,\varphi) &=&
\sum_{\ell=0}^{\infty}\sum_{m=-\ell}^{\ell}a_{\ell m}^TY_{\ell
m}(\theta,\varphi),
\label{imulticoeff} \\
E(\theta,\varphi) &=& \sum_{\ell =2}^{\infty}\sum_{m=-\ell }^{\ell
} \left[\frac{(\ell +2)!}{(\ell -2)!}\right]^{\frac{1}{2}} a_{\ell
m}^EY_{\ell m}(\theta,\varphi),~~~~~~~
\label{emulticoeff} \\
B(\theta,\varphi) &=& \sum_{\ell =2}^{\infty}\sum_{m=-\ell }^{\ell
} \left[\frac{(\ell +2)!}{(\ell -2)!}\right]^{\frac{1}{2}} a_{\ell
m}^BY_{\ell m}(\theta,\varphi),~~~~~~~
\label{bmulticoeff} \\
V(\theta,\varphi) &=& \sum_{\ell =0}^{\infty}\sum_{m=-\ell }^{\ell
}a_{\ell m}^VY_{\ell m}(\theta,\varphi)\label{vmulticoeff}.
\end{eqnarray}

It is important to stress that these expansions are consistent
with the similar definitions in the literature
\cite{kamionkowski1997}, \cite{zaldarriaga1997}.

We are now in position to write down the $I$, $E$ and $B$
functions (\ref{imulticoeff}), (\ref{emulticoeff}) and
(\ref{bmulticoeff}) in terms of the functions
$\alpha(\mu,\varphi)$ and $\beta(\mu,\varphi)$ introduced in
(\ref{vectorn}). From (\ref{stokespar}) we obtain for a
monochromatic radiation beam

\begin{eqnarray}\label{stokesparexp}
I(\eta,\nu,\theta,\varphi) &=&
\frac{h\nu^{3}}{c^{2}}\left[n_{1}(\eta,\nu,\theta,\varphi)+n_{2}(\eta,\nu,\theta,\varphi)\right],~~~~~~
\nonumber \\
Q(\eta,\nu,\theta,\varphi) &=&
\frac{h\nu^{3}}{c^{2}}\left[(n_{1}(\eta,\nu,\theta,\varphi)-n_{2}(\eta,\nu,\theta,\varphi)\right],~~~~~~~~
\nonumber \\
U(\eta,\nu,\theta,\varphi) &=&
-4\frac{h\nu^{3}}{c^{2}}n_{3}(\eta,\nu,\theta,\varphi),
\end{eqnarray}
so that from (\ref{basiscirc}), (\ref{vectorn}) and
(\ref{stokesparexp}) we get (restoring the $\mathbf{n}$-dependence
of the Fourier expansion),
\begin{eqnarray}
I_{\mathbf{n},r}(\eta,\nu,\theta,\varphi) &=&
\frac{h\nu^{3}}{c^{2}}[n^{(0)}(\nu)
+f(\nu){\alpha}_{\mathbf{n},r}(\eta,\mu)(1-\mu^{2})e^{\pm
2i\varphi}],
\label{iintermsn1}\\
Q_{\mathbf{n},r}(\eta,\nu,\theta,\varphi) &=&
\frac{h\nu^{3}}{c^{2}}f(\nu){\beta}_{\mathbf{n},r}(\eta,\mu)(1+\mu^{2})e^{\pm
2i\varphi},
\label{qintermsn1}\\
U_{\mathbf{n},r}(\eta,\nu,\theta,\varphi) &=& \mp
2\frac{h\nu^{3}}{c^{2}}f(\nu){\beta}_{\mathbf{n},r}(\eta,\mu)\mu
e^{\pm 2i\varphi}\label{uintermsn1}.
\end{eqnarray}

From equations (\ref{iintermsn1}-\ref{uintermsn1}) we may readily
evaluate the expressions for $I$, $E$ and $B$, using
(\ref{spherestfpoltensor}), (\ref{metriceps2sphere}),
(\ref{defebmodes}) and (\ref{iintermsn1}-\ref{uintermsn1}); then,
integrating over photon frequencies, we obtain

\begin{eqnarray}\label{stokespar2}
I_{n,r}\left(\mu,\varphi\right)&=& \gamma\left[
\left(1-\mu^{2}\right)\alpha_{n,r}\left(\eta,\mu\right)e^{\pm
2i\varphi}\right], \nonumber \\ E_{n,r}\left(\mu,\varphi\right)&=&
-\gamma \left[\left(1-\mu^{2}\right)
\left(\left(1+\mu^{2}\right)\frac{d^{2}}{d\mu^{2}}+8\mu\frac{d}{d\mu}+12\right)\beta_{n,r}\left(\eta,\mu\right)e^{\pm
2i\varphi}\right], \nonumber \\ B_{n,r}\left(\mu,\varphi\right)&=&
\mp\gamma \left[2\left(1-\mu^{2}\right)
\left(i\mu\frac{d^{2}}{d\mu^{2}}+4i\frac{d}{d\mu}\right)\beta_{n,r}\left(\eta,\mu\right)e^{\pm
2i\varphi}\right],
\end{eqnarray}
where we have defined
\begin{equation}
\label{defgamma} \gamma = \int d\nu~\frac{h\nu^{3}}{c^{2}}f(\nu).
\end{equation}

Once we have the key expressions for evaluating the power spectrum
correlation function all we must do now is solving the Boltzmann
equations (\ref{boltzeq1}) and (\ref{boltzeq2}), which we handle
in the next section.

\section{\label{sec:six} The solutions to the Boltzmann equations}

In the paper \cite{baskaran2006} the authors discuss an analytical
method for solving the Volterra equation represented by
(\ref{boltzeq1}) in terms of a series expansion, and compare their
results with the exact numerical solutions. Here we follow only
their numerical approach, which we sketch below. To do so, we
introduce first the functions
\begin{eqnarray}
\Phi(\eta)=\frac{3}{16}g(\eta)\cal{I}(\eta), \label{phifunct}
\end{eqnarray}
\begin{eqnarray}
H(\eta) = e^{-\tau(\eta)}\frac{dh(\eta) }{d\eta}, \label{hfunct}
\end{eqnarray}
where the function $\tau(\eta)$ represents the \emph{optical
depth} of the universe, and is defined within a time interval
$\eta'$ and $\eta$:
\begin{equation}
\tau(\eta,\eta') = \int_{\eta'}^{\eta}d\eta''q(\eta''); \nonumber
\end{equation}
$g(\eta)$ is the \emph{visibility function}, written as
\begin{equation}
\label{visibfunct}
g(\eta) = q(\eta)e^{-\tau(\eta)} = \frac{d}{d\eta}e^{-\tau(\eta)}.
\end{equation}

Taking $\eta'=\eta_{0}$, we further write the optical depth from a
given conformal instant $\eta$ to the present as
$\tau(\eta_0,\eta)=\tau(\eta)$, that is
\begin{equation}
\tau(\eta) = \int_{\eta}^{\eta_0}d\eta'q(\eta').
\end{equation}

Now, using these definitions, the formal solutions to the
equations (\ref{boltzeq1}) and (\ref{boltzeq2}) are given by the
integral relations \cite{baskaran2006}
\begin{eqnarray}
\beta(\eta,\mu)= e^{\tau(\eta)-in\mu\eta}\int^{\eta}_{0}
d\eta'~\Phi(\eta')e^{in\mu \eta'},
\label{solbetafunct} \\
\xi(\eta,\mu)=e^{\tau(\eta)-in\mu \eta}\int^{\eta}_{0}
d\eta'~H(\eta')e^{in\mu \eta'} \label{solxifunct}.
\end{eqnarray}

Now, since the function $H(\eta)$ is known, we can obtain a single
integral equation for the function (\ref{phifunct}) by plugging
(\ref{solbetafunct}) and (\ref{solxifunct}) into (\ref{defi}), so
that
\begin{equation}
\mathcal{I}(\eta) = e^{\tau(\eta)}\int_{-1}^{1}\int_{0}^{\eta}d\mu
d\eta'\bigg\{ \left(1+\mu^{2}\right)^{2}\Phi(\eta')-
\frac{1}{2}\left(1-\mu^{2}\right)^{2}H(\eta')
\bigg\}e^{in\mu(\eta'-\eta)}; \label{eqfori}
\end{equation}
such expression can be further simplified by introducing the
kernels $K_{\pm}(\eta-\eta')$, defined as
\begin{eqnarray}
K_{\pm}(\eta-\eta')=\int^{1}_{-1}d\mu
(1\pm\mu^{2})^{2}e^{in\mu(\eta-\eta')}, \label{defkernel}
\end{eqnarray}
so that expression (\ref{eqfori}) yields
\begin{equation}
\mathcal{I}(\eta) = e^{\tau(\eta)}\int_{0}^{\eta} d\eta'
\bigg\{K_+(\eta-\eta')\Phi(\eta')-
\frac{1}{2}K_-(\eta-\eta')H(\eta') \bigg\}. \label{eqfori1}
\end{equation}
The final equation for $\Phi(\eta)$ is obtained by multiplying
both sides of this equality by $(3/16)q(\eta)e^{-\tau(\eta)}$ and
using the expression (\ref{phifunct}), so that
\begin{eqnarray}
\Phi(\eta)= \frac{3}{16}q(\eta)\int ^{\eta}_{0}
d\eta'\Phi(\eta')K_{+}(\eta-\eta') + G(\eta), \label{eqforphi}
\end{eqnarray}
where $G(\eta)$ is related to the function (\ref{hfunct}),
\begin{eqnarray}
G(\eta)=-\frac{3}{32}q(\eta)\int^{\eta}_{0}d\eta'
H(\eta')K_{-}(\eta-\eta'). \label{deffunctf}
\end{eqnarray}

The solution to Volterra integral equation (\ref{eqforphi})
provides the values of the functions $\alpha$ and $\beta$ for
every conformal instant $\eta$; in particular, to the present-day
$\eta_{0}$, the expressions $\alpha(\eta_{0},\mu)=\alpha(\mu)$ and
$\beta(\eta_{0},\mu)=\beta(\mu)$ are respectively given by
\begin{equation}\label{alphabetazero}
\alpha_{n,r}(\mu)= \int^{\eta_0}_{0}d\eta\left(H_{n,r}(\eta) -
\Phi_{n,r}(\eta) \right)e^{-i\mu\zeta},~~~~\beta_{n,r}(\mu) =
\int^{\eta_0}_{0}d\eta~\Phi_{n,r}(\eta)e^{-i\mu \zeta},
\end{equation}
where we have introduced the variable $\zeta = n(\eta_0 - \eta)$.
From (\ref{alphabetazero}) we compute the coefficients
$a^{X}_{\ell m}$: to do that we substitute expressions
(\ref{imulticoeff}-\ref{bmulticoeff}) and (\ref{alphabetazero})
into (\ref{stokespar2}), and integrate over angular variables, so
that
\begin{eqnarray}\label{expralm}
a^T_{{\ell} m,nr}&=& (-i)^{\ell-2} \left(\delta_{2,m}\delta_{1,r}
+\delta_{-2,m}\delta_{2,r}\right)a^T_{\ell,nr},
\nonumber \\
a^E_{{\ell} m,nr}&=& (-i)^{\ell-2} \left(\delta_{2,m}\delta_{1,r}
+\delta_{-2,m}\delta_{2,r}\right)a^E_{\ell,nr},
\nonumber \\
a^B_{{\ell} m,nr}&=&(-i)^{\ell-2}
\left(\delta_{2,m}\delta_{1,s}\frac{}{}
-\delta_{-2,m}\delta_{2,s}\right)a^B_{\ell,nr},~~~~~~~
\end{eqnarray}
where
\begin{eqnarray}\label{defat}
a^T_{\ell,nr}&=&{\gamma}\sqrt{4\pi(2{\ell}+1)}
\int_{0}^{\eta_{0}}d\eta~\left(H_{n,r}(\eta) - \Phi_{n,r}(\eta)
\right) T_{{\ell}}(\zeta),\\ \label{defae}
a^E_{\ell,nr}&=&{\gamma}\sqrt{4\pi(2{\ell}+1)}
\int_{0}^{\eta_{0}}d\eta~\Phi_{n,r}(\eta) E_{{\ell}}(\zeta),\\
\label{defab} a^B_{\ell,nr}&=&{\gamma}\sqrt{4\pi(2{\ell}+1)}
\int_{0}^{\eta_0}d\eta~\Phi_{n,r}(\eta) B_{{\ell}}(\zeta),
\end{eqnarray}
and $T_{\ell}(\zeta)$, $E_{\ell}(\zeta)$, $B_{\ell}(\zeta)$ are
the multipole projection functions which appear after the
integration over the angular variables, whose form are given by
\begin{eqnarray}\label{multprojfunct}
T_{{\ell}}(\zeta)&=&
\sqrt{\frac{\left(\ell+2\right)!}{\left(\ell-2\right)!}}\frac{j_{\ell}(\zeta)}{\zeta^{2}},\nonumber \\
E_{{\ell}}(\zeta)&=&
\left[\left(2-\frac{l(l-1)}{\zeta^{2}}\right)j_{\ell}(\zeta)-\frac{2}{\zeta}j_{{\ell}-1}
(\zeta)\right],\nonumber \\
B_{{\ell}}(\zeta)&=&
2\left[-\frac{({\ell}-1)}{\zeta}j_{\ell}(\zeta)+j_{{\ell}-1}(\zeta)\right].
\end{eqnarray}

Once we have obtained the expressions for (\ref{expralm}) we can evaluate the most important tool of CMB physics: the correlation function
$C_{\ell}^{XX'}$, where $X,X'=E,B$. Since each $a^{X}_{\ell m}$ in (\ref{emulticoeff}), (\ref{bmulticoeff}),
depends upon the wavenumber $n$ and polarization
state $r$, we write $a^{X}_{\ell m}$ as $a^{X}_{\ell m,nr}$, so that the correlation function is given by \cite{baskaran2006}:
\begin{equation}
C_{\ell}^{XX'} = \frac{{\cal{C}}^{2}}{4\pi^{2}(2\ell+1)}\int ~ndn
\sum_{r=1,2}\sum_{m=-\ell}^{\ell} [ a^{X}_{\ell m,nr}a^{X'*}_{\ell
m,nr}+ a^{X*}_{\ell m,nr}a^{X'}_{\ell m,nr} ].
\label{correlationcoeff}
\end{equation}

To conclude this section, let us say some words about the ionization
history of the universe. We focus here on the time of decoupling,
were the CMB radiation was released from the primordial nuclei. At
decoupling, which took place around redshift $z\sim 1088$, the
universe underwent a transition from a completely ionized state to
a state in which neutral hydrogen and helium atoms were formed. In
this process the radiation decoupled from the matter, originating
the CMB radiation and a neutral pre-galactic baryonic medium (PGM). Then, at some redshift between $14<z<6$ the PGM was ionized again by the
UV radiation from the first luminous objects, leaving the intergalactic medium (IGM) ionized \cite{Fan:2006dp}. Such process is called
\emph{reionization}, and would leave observable imprints on the CMB polarization spectrum due to the interactions
of the CMB photons with the free electrons now available due to the reionized medium \cite{Bond:1984fp}, \cite{Vittorio:1997vt}.
However, the reionization epoch is still not fully understood, and many models have been proposed to shed a light on the physics of this process
(see \cite{Lee:2009wa} and references therein), which can be homogeneous models with a sudden reionization (\emph{e. g.} as discussed in
\cite{Xia:2009qd}, \cite{Giannantonio:2007za}), or extended models with double reionization \cite{Cen:2002zc}, among others
(see \cite{Xia:2009qd} for a more comprehensive list of papers).

In the present paper we shall consider solely the epoch of
recombination, whose physical process is very well understood.
Despite reionization is fundamental to understand the
low-multipole behavior of CMB polarization, it can be neglected in
a first-approximation to study temperature anisotropies generated
by the tensor modes. Next, as for the decoupling epoch, the
formulae for the density of free electrons $N_{e}(\eta)$ and the
fraction of ionized electrons are given by \cite{peebles1993}
\begin{eqnarray}
N_e(\eta) = \left(1-\frac{Y_p}{2}\right)\frac{X_e(\eta) \Omega_b
\rho_c}{m_p}\left(\frac{a(\eta_0)}{a(\eta)}\right)^{3},\nonumber
\end{eqnarray}
and \cite{hu1995}
\begin{equation}
X_e(\eta) =
\left(1-\frac{Y_p}{2}\right)^{-1}\left(\frac{c_2}{1000}\right)
\left(\frac{m_p}{2\sigma_T R_H \rho_c}\right)\Omega_b^{c_1-1}
\left(\frac{z}{1000}\right)^{c_2-1}
\left(\frac{a'}{a}\right)(1+z)^{-1}. \label{fracionelec}
\end{equation}
In these formulae $Y_p \approx 0.23$ is the primordial helium mass
fraction, $\Omega_b$ is the baryon content, and $m_p$ is the mass
of a proton. The constants are given by $c_1 = 0.43$, $c_2 =
16+1.8\ln\Omega_B$; we take $\Omega_b=0.046$ \cite{spergel2007}.

\section{\label{sec:seven} CMB Anisotropies in Massive Gravity}

Let us now compute the Boltzmann equations for radiative transfer
in the presence of weak gravitational fields in Massive Gravity.
The anisotropy and polarization power spectra, as discussed in
sections \ref{sec:four}, \ref{sec:five} and \ref{sec:six}, is a
combination of two physical processes appearing in the Boltzmann
equation (\ref{boltz1}): Thomson scattering, represented by the
collisional integral (\ref{collterm}) on its right-hand side, and
the gravitational redshift of the photon, represented by the
geodesic equation (\ref{geodgen}) on its left-hand side. The
collisional integral for Massive Gravity is the same as in GR,
since the process of scattering involves only photons and
electrons, and the lagrangian of matter is minimally coupled to
the metric. The basic change is due to the different gravitational
field strength represented by $h_{ij}$ in the geodesic equation,
which now depends on the details of the underlying gravitational
model. However, the general form of Boltzmann equations
(\ref{boltzeq1}) and (\ref{boltzeq2}) are the same for both GR and
Massive Gravity (see \cite{bessada2009} for details), which we
proceed to evaluate numerically in what follows.

We start our analysis of the possible signatures of massive
gravitons on CMB anisotropies in the light of our discussion in
section \ref{primgwmassive} and the results reviewed in section
\ref{sec:six}. For the sake of simplicity we consider first the
case of GR, and then extend our discussion to Massive Gravity.
From figure \ref{fig1} we see that long wavelength tensor modes
(that is, low $n$, in the figure corresponding to $n=10$) remain
frozen during decoupling, so that they won't contribute
significantly to CMB anisotropies and polarization at this epoch;
this effect will be mainly due to the short wavelength tensor
modes (high $n$, in the figure corresponding to $n=10^2$ and
$n=10^3$; notice that tensor modes with $n=10^4$ are ``dead" at
decoupling), which fell into the horizon at an earlier time. For
late times the situation is inverse: long wavelength modes are now
dead, and short wavelength modes are falling into the horizon.
Therefore, this quick and simple analysis leads us to conclude
that short wavelength tensor modes will be predominant over early
times, and long wavelength modes at late times. In terms of
multipoles of the radiation this feature can be understood in the
following way: short wavelength modes and early times correspond
to larger values of the variable $\zeta = n(\eta_0 - \eta)$, so
that the spherical Bessel functions appearing in
(\ref{multprojfunct}) will have nonzero values only for higher
multipoles; conversely, long wavelengths and late times will give
rise to smaller $\zeta$, and hence $j_\ell(\zeta)$ will be nonzero
only for low multipoles.

Therefore, roughly speaking, in GR long wavelength tensor modes
will influence mainly the CMB low multipoles, whereas short
wavelength tensor modes will substantially contribute to the
polarization on higher multipoles. This is also true in the case
of Massive Gravity, but now with one substantial difference: long
wavelength modes fall into horizon earlier than their massless
counterparts, as seen in figures \ref{fig2} and \ref{fig3},
altering then the form of the signature at low multipoles. Also,
the oscillations noticed in the late-time evolution of the massive
modes, valid not only for those possessing long wavelengths, but
for the ones possessing short wavelengths as well (as seen in
figure \ref{fig8}, for instance) would contribute to yield a
distinct signature on low multipoles.

However, since we are not considering reionization in this paper,
the distinct signatures discussed above will not show up on the
polarization spectrum. This happens because in this case the
visibility function is zero at this epoch, and then the source
functions $\Phi$ (\ref{phifunct}) will be zero, since they depend
linearly on $g(\eta)$. However, this fact does not affect CMB
anisotropies, since the mode coefficients $a^T_{\ell m}$, given by
(\ref{defat}), depends on $H(\eta)$, defined by (\ref{hfunct}),
which is not zero even in the absence of reionization. Since the
source function $H(\eta)$ depends on the tensor mode amplitudes,
and massive and massless modes are different at late times, we
conclude that the massive modes would leave a distinct signature
on CMB low multipoles even without reionization.

Let us now perform a numerical analysis in order to check the
points discussed above. As we have discussed in Section
\ref{primgwmassive}, if the mass of the tensor mode is less or
equal than $m_l$, Massive Gravity produces the same results as GR;
we choose then masses within the range $m= 10^{-27}cm^{-1}$ -
$m=10^{-26}cm^{-1}$, whose associated anisotropies power spectrum
is depicted in figure \ref{ttpower}. Figure \ref{ttpowerzoom} show
the low-multipole region of the correlation function \ref{ttpower}
in detail.

\smallskip
\FIGURE{\epsfig{file=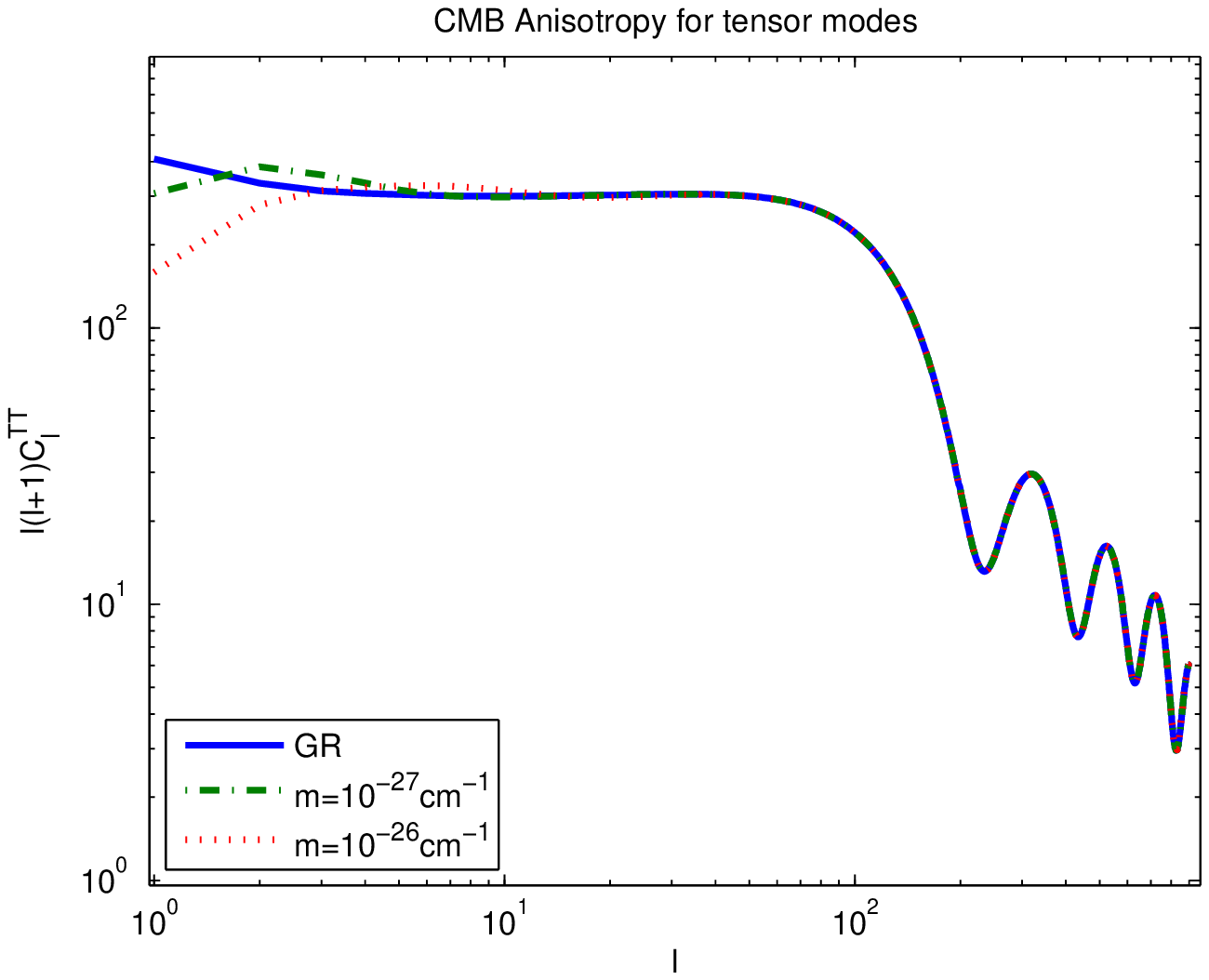,width=9cm}
        \caption[]{The correlation
functions $C^{TT}_{\ell}$ for GR and Massive Gravity. Notice that
the massive gravitons leave a signature on the spectrum for low
multipoles.}%
    \label{ttpower}}

This figure show distinct signatures for massless and massive
gravitons, as we have argued above. Therefore, for the range of
masses selected, massive tensor modes leave a clear signature on
low multipoles $\ell<30$. Since the heavier modes fall into the
horizon earlier, they have the stronger signature, as shown. If we
had chosen a different mass, say $m=10^{-21}cm^{-1}$, the
signature would be stronger, and possibly would appear for
multipoles $\ell>30$. This can be explained by simply analyzing
the trend shown in figures \ref{fig2} and \ref{fig3}: the heavier
the mass, the earlier the modes fall into the horizon, which
correspond to higher multipoles. However, even in this case, as
the trend shown in figure \ref{ttpowerzoom} indicates, the
signature will be particularly strong on low multipoles.
Therefore, if the tensor modes of the metric fluctuations are
massive, they could be detected directly by the CMB anisotropy
power spectrum if their mass are greater than the limit $m_l\sim
10^{-29}cm^{-1}$, and their signatures would be noticeable
specially on low multipoles.
\FIGURE{\epsfig{file=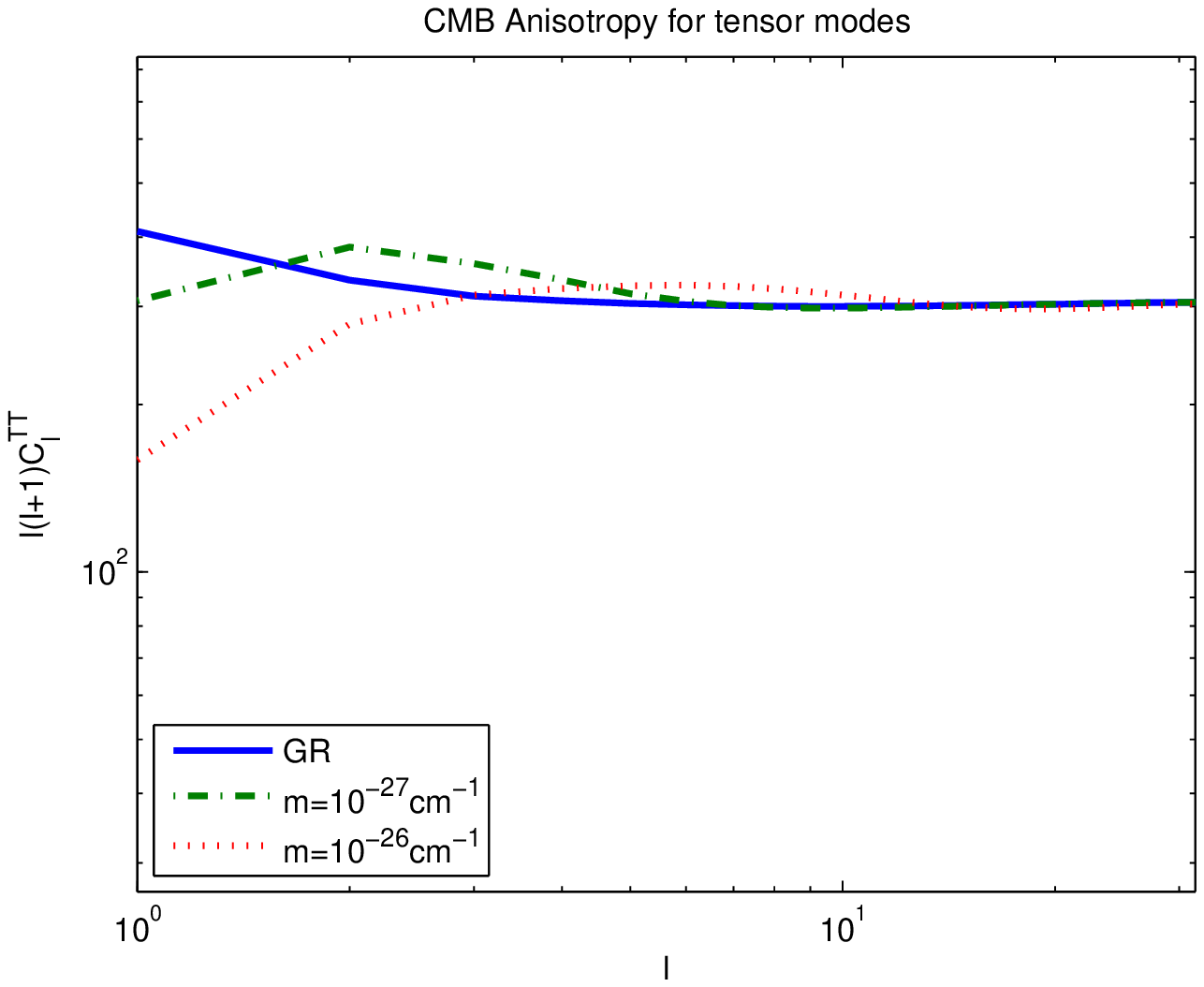,width=10cm}
        \caption[]{The
low-multipole ``tail" in the TT correlation function. Notice the
quite distinct signatures for $\ell < 30$ for the mass range selected.}%
    \label{ttpowerzoom}}

Therefore, the results above indicate clearly that the future
measurements on the TT correlation might be decisive for probing
the existence of massive tensor modes, for the signature left by
them could be strong enough to be distinguished from those of the
massless modes.

\section{\label{conc}Conclusions}

In this work we have first studied the time evolution of massive
tensor modes and shown that there is a graviton mass limit, $m_l\sim 10^{-29}cm^{-1}$, such that gravitons with
masses $m\leq m_l$ behave indistinguishably from
massless gravitons. The same happens to gravitons with short wavelengths
(wavenumbers $n\geq 100$ in our example): their behavior is
almost the same as of the massless gravitons for all the masses taken
into account here.

We have also shown that long wavelength massive tensor modes fall
into the horizon earlier their massless counterparts, whereas
short wavelength modes behaves quite similarly as in GR. The net
effect of this behavior, as we have shown in the TT correlation
function plotted in figures \ref{ttpower} and \ref{ttpowerzoom},
is a distinguished signature on low multipoles; the heavier the
mass of the mode, the stronger is its signature compared to that
of massless gravitons. For the range of masses considered here,
$m= 10^{-27}cm^{-1}$ - $m=10^{-26}cm^{-1}$, the signatures show up
at $\ell<30$; however, we have argued that such signatures might
appear at $\ell>30$ in the case of masses greater than
$m=10^{-25}cm^{-1}$.

Therefore, our results indicate that the future precise
measurements of the CMB anisotropies induced by tensor modes might
be decisive for probing the existence of massive gravitons, for
the signature left by them could be strong enough to be
distinguished from those of the massless modes.

\acknowledgments DB and ODM thank Odylio D. Aguiar, Jos\'e Carlos
N. de Araujo, Armando Bernui, Thyrso Villela and Carlos Alexandre
Wuensche for very helpful discussions. The authors also thank
Jos\'e A. de Freitas Pacheco for very important discussions and
for a critical reading of the manuscript. The authors would like
to thank the referee for helpful comments that we feel
considerably improved the paper. DB thanks Cesar A. Costa and
Cl\'audio Brand\~ao for a great help on numerical methods. DB was
financially supported by CAPES, and ODM is partially supported by
CNPq (grant 305456/2006-7).

\end{document}